\documentstyle[12pt,epsf]{article}
\columnwidth\textwidth

\begin{document}
\renewcommand{\theequation}{\arabic{equation}}
\parskip=4pt plus 1pt
\textheight=8.7in
\textwidth=6.0in
\newcommand{\be}{\begin{equation}}
\newcommand{\ee}{\end{equation}}
\newcommand{\bea}{\begin{eqnarray}}
\newcommand{\eea}{\end{eqnarray}}
\newcommand{\co}{\; \; ,}
\newcommand{\nn}{\nonumber \\}
\newcommand{\pnn}{\nonumber}
\newcommand{\scs}{\co \;}
\newcommand{\per}{ \; .}
\newcommand{\unith}{{\bf{\mbox{1}}}}
\newcommand{\la}{\langle}
\newcommand{\ra}{\rangle}
\newcommand{\Pc}{\bar{\phi}}
\newcommand{\nonoverline}{}
\newcommand{\mput}[1]{\mbox{\tiny{#1}}}
\newcommand{\beq}{\begin{equation}}
\newcommand{\eeq}{\end{equation}}
\newcommand{\beqa}{\begin{eqnarray}}
\newcommand{\eeqa}{\end{eqnarray}}
\newcommand{\beqan}{\begin{eqnarray*}}
\newcommand{\eeqan}{\end{eqnarray*}}
\newcommand{\ba}{\begin{array}}
\newcommand{\ea}{\end{array}}
\newcommand{\ul}{\underline}
\newcommand{\ol}{\overline}
\newcommand{\olc}{\bar}
\newcommand{\Ra}{\Rightarrow}
\newcommand{\ve}{\varepsilon}
\newcommand{\vp}{\varphi}
\newcommand{\wt}{\widetilde}
\newcommand{\wh}{\widehat}
\newcommand{\cL}{{\cal L}}
\newcommand{\dfrac}{\displaystyle \frac}
\newcommand{\grts}{\stackrel{>}{_\sim}}
\newcommand{\lets}{\stackrel{<}{_\sim}}
\newcommand{\del}{\partial}
\newcommand{\dis}{\displaystyle}
\newcommand{\bla}{\left\langle}
\newcommand{\bra}{\right\rangle}
\newcommand{\no}{\nonumber}
\newcommand{\bdm}{\begin{displaymath}}
\newcommand{\edm}{\end{displaymath}}

\thispagestyle{empty}
\begin{flushright}
LU/TP 97-14\\
UWThPh-1997-15\\
BUTP-97/17\\
HIP-1997-31 /TH\\
\end{flushright}
\vspace{2cm}
\begin{center}
\begin{Large}
Pion--pion scattering at low energy
 \\[1cm]
\end{Large}
J. Bijnens$^1$, G. Colangelo$^{2}$, G. Ecker$^3$, J. Gasser$^4$
and M.E. Sainio$^5$ \\[2cm]
${}^1$ Dept. of Theor. Phys., Univ. Lund, S\"olvegatan 14A, S--22362 Lund\\
${}^2$ INFN -- Laboratori Nazionali di Frascati, P.O. Box 13, I--00044
Frascati \\
${}^3$ Inst. Theor. Phys., Univ. Wien, Boltzmanng. 5, A--1090 Wien\\
${}^4$ Inst. Theor. Phys., Univ. Bern, Sidlerstr. 5, CH--3012
Bern\\
${}^5$ Dept. of Physics, Univ. Helsinki, P.O. Box 9, FIN--00014
Helsinki \\[1cm]
July 1997\\[1cm]
{\bf Pacs:} 11.30Rd, 12.39.Fe, 13.75.Lb\\[0.2mm]
{\bf Keywords:} \begin{minipage}[t]{9.5cm} Chiral Symmetry,
Chiral Perturbation Theory,\\Pion--Pion Scattering\end{minipage}
\end{center}
\begin{abstract}
\noindent
We present technical details of the evaluation of the
elastic $\pi\pi$ scattering amplitude to two loops in chiral perturbation
theory. In particular, we elaborate on
the renormalization procedure at the
two--loop order and on the evaluation of the relevant Feynman
diagrams that can all be expressed in terms of elementary
functions.  For the sake
of clarity, we discuss these matters both in the
$N$--component $\phi^4$ theory
(in its symmetric phase) and in chiral perturbation theory.
Estimates for the relevant low--energy constants of $O(p^6)$ are  presented.
Threshold parameters and phase shifts are then calculated for two
sets of $O(p^4)$ coupling constants and compared with experiment.
We comment on the extraction of threshold parameters from phase shift data.
\end{abstract}
\setcounter{page}{0}

\clearpage
\tableofcontents
\section{Introduction}
Elastic $\pi\pi$ scattering is a fundamental process for QCD at low
energies. It provides an ideal testing ground for the mechanism of
spontaneous chiral symmetry breaking. Since only the (pseudo--)Goldstone
bosons of chiral $SU(2)$ are involved, one expects the low--energy
expansion of the scattering amplitude to converge rather rapidly. The
systematic procedure for this low--energy expansion is called chiral
perturbation theory (CHPT) \cite{wein79,glann,glnp1,hlann}.

In the framework of CHPT,
the $\pi\pi$ scattering amplitude was evaluated at lowest order in
the chiral expansion in \cite{weinpipi} and to next--to--leading order in
\cite{glann,glpl}. The theoretical developments up to 1994 are summarized
in \cite{DAFNE2,MIT}. The forthcoming experimental improvements concerning
$\pi\pi$ scattering are also discussed there. The amplitude to
next--to--next--to--leading order was recently calculated in a dispersive
approach in \cite{KMSF95} and in the standard CHPT framework in \cite{BCEGS1}.
Other developments since 1994 can be found in
Refs.~\cite{wanders97}-\cite{gkms97}.

Both the explicit  calculation of the Feynman diagrams and
the renormalization procedure needed to evaluate the scattering
amplitude  at  the two--loop order are quite involved.
The  main aim of the present article
is to detail the methods used in \cite{BCEGS1} for that purpose.

The  amplitude contains a nontrivial
analytical part that can be expressed in terms of  logarithms
that generate the cuts required by unitarity.
In addition, there is a contribution
which consists of a polynomial in the external momenta.
Our calculation reveals the dependence of the six coefficients in this
polynomial on the pion mass and on the low--energy constants of both
$O(p^4)$ and $O(p^6)$. We find  that the complete
two--loop amplitude can be  expressed in terms of
elementary functions.

We found it useful to first illustrate in Sect. 2
the corresponding calculation in the
$N$--component $\phi^4$ theory
(in its symmetric phase), where the algebraic manipulations needed are simple.
The loop expansion is
presented together with the renormalization at both the one-- and two--loop
levels. A technical
issue that also comes up in the CHPT calculation is the use of the
equations of motion (EOM) for the counterterms. We show how
different forms of the counterterms at the one--loop level  lead to the same
generating functional
after a proper redefinition of the counterterms at the two--loop order.

In Sect.~3, the analogous calculation is performed in CHPT. The
renormalization procedure is again discussed in detail, both for minimal
subtraction and for the modified minimal subtraction that we actually use.
For the effective chiral lagrangian, the role of EOM
is similar as in
$\phi^4$,
albeit algebraically
more involved. In Sect. 4, the results for the pion
mass, the pion decay constant and the $\pi\pi$ scattering amplitude are
presented to $O(p^6)$ in the modified minimal subtraction scheme.
Here, we also compare our results with the calculation of
Ref. \cite{KMSF95}.

In Sect.~5, we perform a numerical analysis to compare with experimental
information. For this purpose, we first derive estimates for the low--energy
constants of $O(p^6)$. We include the effects of vector and scalar meson
resonances and of $K$ and $\eta$ contributions. We then discuss numerically
the sensitivity of the amplitude to those constants. We emphasize that the
main uncertainties are due to the couplings of
$O(p^4)$. We use two sets of these couplings to illustrate the uncertainty:
The first set used in \cite{BCEGS1} is essentially based on phenomenology
to $O(p^4)$ \cite{glann,bcgke4}. For the second set, we use the present
calculation to extract the couplings $\olc{l}{_1}$, $\olc{l}{_2}$ from
the observed
$D$--wave scattering lengths. For the two sets, we then analyse threshold
parameters and phase shifts and compare them with available experimental
results.

Sect.~6 contains our conclusions. Some technical aspects
on the calculation of two--loop diagrams, the off--shell
scattering amplitude of $O(p^4)$ and analytic expressions
for the scattering lengths as well as for the coefficients $b_i$ in the
$O(p^6)$ amplitude are relegated to four appendices.

\renewcommand{\theequation}{\arabic{section}.\arabic{equation}}
\addtocounter{section}{0}
\setcounter{equation}{0}
\section{$N$--component $\phi^4$ to two loops}
 As we have mentioned in the introduction,
we illustrate
the loop expansion in CHPT first with the
$N$--component $\phi^4$ theory
in the unbroken phase. For this
purpose, we proceed in a manner as analogous as possible to the
chiral expansion. In particular, we rely on  the external
field technique and use a procedure that is scale independent
at each step, as is done in CHPT. We also comment on the role
of counterterms that vanish upon use of the EOM
and give  the relation to the MS scheme.
\subsection{The loop expansion}
\begin{figure}[t]
\begin{center}
\mbox{\epsfysize=8cm \epsfbox{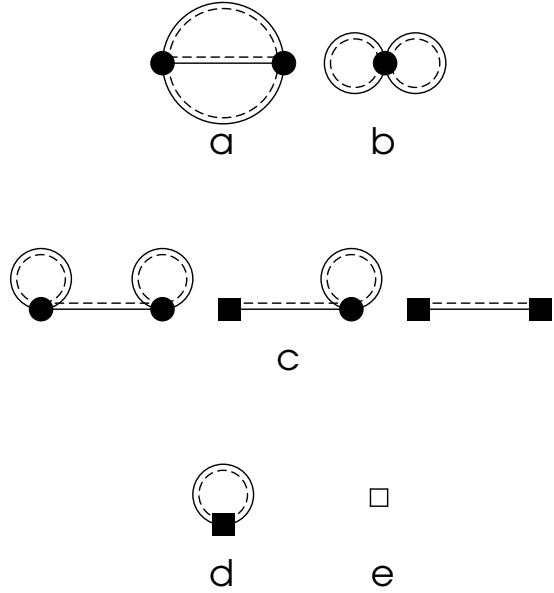}
     }
\caption{The two--loop graphs in $\phi^4$ theory. Graph a
corresponds to the contribution (\protect\ref{eq_za}), and
analogously for the diagrams b--e. The solid--dashed lines stand
for the propagator (\protect\ref{eqpropg}). The black circles
denote
a vertex generated by the lagrangian $\left.{\cal
L}\right|_{C_i=0}$ in (\protect\ref{eqlagrangian}),
and the filled squares stand for the contributions from the
counterterm $C_1$. Finally, the contributions from the
counterterm $C_2$ are indicated with an open square in graph e.}
\label{f1} \end{center} \end{figure}

The loop expansion is equivalent to an expansion of the
generating functional in powers of
$\hbar$ -- we therefore explicitly display $\hbar$ in this
section. The lagrangian is
\bea
{\cal {L}}&=&\frac{1}{2}(\partial_\mu \phi^T\partial^\mu\phi
-M^2\phi^T\phi) -\frac{g}{4}(\phi^T\phi)^2 -\phi^Tf
-\sum_{\nu=1}^{\infty}\hbar^\nu C_\nu\co
\label{eqlagrangian}\eea
where the symbol $\phi$ collects the $N$ fields
 $\phi_1,\ldots ,\phi_N$,
\bea
\phi^T=(\phi_1,\ldots ,\phi_N)\per\pnn
\eea
Below, we will use an analogous notation for  any multicomponent
vector, e.g.,
$a^T=(a_1,a_2,a_3)$, etc.
 The quantities $C_\nu$
denote the counterterms that remove the ultraviolet singularities
at order $\hbar^\nu$. They are linear combinations of
$O(N)$--invariant polynomials  of dimension $\leq$ 4.
Considering the external field $f$  to be a  spurion $O(N)$
vector
and using partial integration to eliminate $\partial_\mu\phi^T
\partial^\mu\phi$, one is left with
 \bea
P_1&=&\frac{1}{2}\phi^T\Box\phi\scs
P_2=\frac{1}{2}M^2\phi^T\phi\scs
P_3=\frac{g}{4}(\phi^T\phi)^2\scs\label{eqp_i}
P_4=\phi^T f\per
\eea
As is shown below, one may eliminate one of these
polynomials by use of the EOM.
 Here we discard $P_4$ and write
\bea
 C_1&=&ga^TP\scs C_2=g^2b^TP\scs\ldots\scs\nn\label{eqc_i}
P^T&=&(P_1,P_2,P_3)\per
\eea
 In the following
we use dimensional regularization. As a result,
 the vectors $a$ and $b$  become $M$--independent
functions of the space--time dimension $d$, divergent as
$d\rightarrow 4$. The external field $f$ allows one
to construct the generating functional by evaluating the
vacuum--to--vacuum transition amplitude
 \bea
\exp \{iZ[f]/\hbar\}=N\int [d\phi] \exp\{iS/\hbar\}\scs
S=\int dx\, {\cal L}\per
\eea
The normalization constant $N$ ensures that $Z[0] =0$.
 Expanding the
right--hand side in powers of $\hbar$ generates the series
\bea
Z&=&Z_0+\hbar Z_1 +\hbar^2 Z_2 +O(\hbar^3)\per\pnn
\eea
To arrive at the explicit expressions for the components $Z_i$,
one considers fluctuations
around the classical solution $\Pc$ that is determined by
the external field through the EOM,
\bea
(M^2 + \Box)\Pc^a+g\Pc^T
{\bar{\phi}}{\bar{\phi}}^a+f^a=0\;\; ; \; \; a=1,\ldots,N\per
\label{eqeom} \eea
Below,  barred quantities always denote
quantities evaluated at (\ref{eqeom}), e.g.,
\bea
 \bar{S}=\int dx\, {\cal L}(\Pc[f],f)\scs
\eea
etc.
We write the
fluctuations in the form \bea \phi=\Pc +\xi\co
\eea
use  translation invariance of the measure,
 $[d\phi]=[d\xi]$, and find
\bea
\exp\{iZ/\hbar\}&=&N\exp\{i
\bar{S}/\hbar\}\times
 \nn
 &&\int[d\xi]\exp\{\frac{-i}{\hbar}\int
dx\,[\frac{1}{2}\xi^T D\xi +F(\xi,\Pc)]\}\per
\eea
The
differential operator $D$ acts in coordinate and flavour space,
\bea
D^{ab}&=&D_0^{ab}+\sigma^{ab}\scs\nn
D_0^{ab}&=&(M^2+\Box)\delta^{ab}\scs
\sigma^{ab}=g\{\Pc^T\Pc\delta^{ab}+2\Pc^a\Pc^b\}\per
\eea
Furthermore, considering the fluctuations to be of order
$\hbar^{1/2}$, the quantity $F$ starts at order $\hbar^{3/2}$.
For the evaluation of the generating functional at two--loop
order, it is sufficient to keep terms of  order $\hbar^2$ in
$F$,
\bea
F&=&g\xi^T\xi\xi^T\Pc +\frac{g}{4}(\xi^T\xi)^2+
\hbar g\{\xi^Td_1\Pc+\xi^Td_2\xi\}
+O(\hbar^{5/2})\co\nn
d_1^{ab}&=&(a_1\Box +a_2M^2+a_3g\Pc^T\Pc)\delta^{ab}\co\nn
d_2^{ab}&=&\frac{1}{2}\left(a_1\Box+a_2M^2\right)\delta^{ab}+
\frac{a_3}{2}\sigma^{ab}\per\eea
We then arrive at the following  expressions
for the components $Z_i$,
\bea \label{eqz_0}
Z_0&=&-\frac{1}{2}\int dx\,\left\{
\ol{P}_4-2\ol{P}_3\right\}\co\\
 Z_1&=&\frac{1}{2}\int dx\,\left\{i\langle
\ln(1+D_0^{-1}\sigma)\rangle-
2\overline{C}_1\right\}\label{eqz_1} \co \eea
\newcounter{zahler1}
\addtocounter{zahler1}{1}
\renewcommand{\theequation}{\arabic{section}.\arabic{equation}
\alph{zahler1}}
\vskip-10mm
\bea
Z_2&=&-g^2\int dx dy\, \Pc_x^T\left\{2G^3_{xy}+G_{xy}\langle
G^2_{xy}\rangle\right\}\Pc_y\label{eq_za}\\
\addtocounter{zahler1}{1}\addtocounter{equation}{-1}
&&+\frac{g}{4}\int dx\, \left\{\left[\langle
G_{xx}\rangle^2+2\langle G^2_{xx}\rangle\right]
-\left[\cdots\right]_{\Pc=0}\right\}\\
\addtocounter{zahler1}{1}\addtocounter{equation}{-1}
&&-\frac{g^2}{2}\int dx dy \,\Pc^T_x\left\{\langle
G_{xx}\rangle
+2G_{xx}+i{d_{1x}}\right\}G_{xy}\times\nn
&&\hspace{3cm}\left\{\langle G_{yy}\rangle +2G_{yy}
+i{d_{1y}}\right\}\Pc_y \\
\addtocounter{zahler1}{1}\addtocounter{equation}{-1}
 &&+ig\int dx\,
\left\{\langle d_{2y}G_{yx}\rangle_{y=x} -\langle
\cdots\rangle_{y=x,\Pc=0} \right\}\\
\addtocounter{zahler1}{1}\addtocounter{equation}{-1}
&&-\int dx\,\overline{C}_2\per\label{eqz_2}
 \eea
\renewcommand{\theequation}{\arabic{section}.\arabic{equation}}
Here we have introduced the propagator in the presence of the
external field $f$,
\bea
G_{xy}&=&\int_0^\infty d\lambda \langle x|e^{-\lambda
D}|y\rangle\scs\nn
D^{ab}_xG^{bc}_{xy}&=&\delta^{ac}\delta^{(d)}(x-y)\per
\label{eqpropg}
\eea
The symbol $\langle A \rangle$
 denotes the trace
 of the matrix $A$ in flavour space, and
$\Pc_x=\bar{\phi}(x)$.
 The various contributions to $Z_2$ are illustrated
in  Fig. \ref{f1}.

The numbering is such that the
contribution (\ref{eq_za}) corresponds to Fig. \ref{f1}a, etc.
We refer in the following to Fig. \ref{f1}a (b) as the
 sunset (butterfly) diagram, respectively.
Below, we will also need the free  propagator,
\bea
\triangle_{xy}&=&G_{xy}|_{f=0}\scs
\eea
which is a diagonal $N\times N$ matrix in flavour space.

 \subsection{Renormalization}
The above expressions have a well--defined limit as the
space--time dimension $d$ reaches its physical value $d=4$ only for an
appropriately chosen singular behaviour of the vectors $a$ and
$b$ as $d\rightarrow 4$. In
this subsection, we display and comment these
singularities and start the discussion with the one--loop graphs.

\subsubsection{One--loop diagrams}
The singularities at one--loop order may easily be
identified by writing
\bea
\langle\ln(1+D_0^{-1}\sigma)\rangle=
\langle
D_0^{-1}\sigma\rangle-\frac{1}{2}\langle D_0^{-1
}\sigma D_0^{-1}\sigma\rangle+\cdots\co\pnn
\eea
where the terms omitted are finite at $d=4$. Evaluating the trace
and keeping only the singular terms at $d=4$ gives
 \bea
i\int\! dx\,\langle
\ln(1+D_0^{-1}\sigma)\rangle = -g(N+2)T_M\!\int\! dx\, \Pc^T\Pc
-\frac{g^2}{2}(N+8)\dot{T}_M\!\int\! dx\, (\Pc^T\Pc)^2
+ \cdots , \hspace{-1cm}\nn
\pnn\eea
where  the tadpole integral $iT_M$ is the free propagator at
coinciding arguments in coordinate space,
 \bea\label{eqtad}
T_M=\frac{1}{i}\int_0^\infty d\lambda \langle x|e^{-\lambda
D_0}|x\rangle
=\frac{M^{d-2}}{(4\pi)^{d/2}}\Gamma(1-d/2)\co
\eea
and
\bea\label{eqtaddot}
\dot{T}_M\doteq \partial_{M^2}T_M=
-\frac{M^{d-4}}{(4\pi)^{d/2}}\Gamma(2-d/2)
\per
\eea
To determine $C_1$, we work out the
singular part in the  integrals $T_M$ and $\dot{T}_M$. We
introduce the scale $\mu$, \bea
M^{2w}=\mu^{2w} (M/\mu)^{2w}\scs w=d/2-2\co\nonumber
\eea
and obtain
\bea\label{eqtadp}
T_M&=&\frac{M^2}{16\pi^2}\mu^{2w}\left\{\frac{1}{w}+a(M/\mu)-1+
b(M/\mu)w+O(w^2)\right\}\co\nn
\dot{T}_M&=&\frac{\mu^{2w}}{16\pi^2}\left\{\frac{1}{w}+a(M/\mu)+c
( M / \mu)w+O(w^2)\right\}\co
\eea
with
\bea\label{eqtada}
a(M/\mu)=\ln(M^2/\mu^2)-\Gamma'(1)-\ln(4\pi)\per
\eea
The functions $b(M/\mu)$ and $c(M/\mu)$ will not be needed in
explicit form in
the following. Since we have introduced the scale $\mu$ in such a
manner
that $T_M$ and $\dot{T}_M$ are scale--independent, we may use
 scale--independent counterterms as well,
\bea
a_1&=&0\co\nn
a_2&=&\frac{\mu^{2w}}{16\pi^2}\left\{\frac{-(N+2)}{w}
+a_2^r(\mu,w)\right\}\co\nn
a_3&=&\frac{\mu^{2w}}{16\pi^2}\left\{\frac{-(N+8)}{w}+
a_3^r(\mu,w) \right\}\co\label{eqa_i}
\eea
where the renormalized couplings $a_i^r(\mu,w)$ are finite at
$w=0$, with
  \bea\label{eqasc} \left(\mu
\frac{\partial}{\partial\mu}+2w\right)a_2^r(\mu,w)=2(N+2)\co
\eea
and analogously for $a_3$.
 With the choice  (\ref{eqa_i}, \ref{eqasc}),
the one--loop functional $Z_1$ is scale independent
and  finite at $d=4$.

\subsubsection{Two--loop diagrams}
In the next step, we determine the counterterms that
render the two--loop contributions finite at $d=4$. For this
purpose, we identify the singular part of the full propagator
 \cite{JO82},
\bea
S_{xy}&=&\triangle_{xy}+\sigma_x\dot{\triangle}_{xy}\scs\nn
\dot{\triangle}_{xy}&=&\partial_{M^2}\triangle_{xy}\per\pnn
\eea
The remainder $R$,
\bea
G_{xy}&=&S_{xy}+R_{xy}\co\label{eqsingreg}
\eea
is finite at $x=y$. The
decomposition (\ref{eqsingreg}) generates in
 the two--loop functional $Z_2$
terms of the form $R^3,R^2 S, R S^2$ and $S^3$.
 A
closer examination
reveals that all nonlocal singularities proportional to
$R$ and to $R^2$
cancel out after putting in the values of the $a_i$ from Eq.
 (\ref{eqa_i}). The divergent  contribution to  $Z_2$ is
therefore obtained
 by replacing everywhere in $Z_2$ the propagator $G$
by its singular part, $G\rightarrow S$. We now sketch how the
singular pieces are evaluated, and consider for illustration the
part proportional to $G_{xy}^3$ in Eq. (\ref{eq_za}) that
contributes to the sunset diagram Fig. \ref{f1}a.
 After $G \rightarrow S$,
we expand the
field $\Pc_y$  around $y=x$
 and are then  left with divergent
vacuum integrals of the form
\bea
\int dx \left(\triangle^3_{xy};\,
 \triangle^2_{xy} \dot{\triangle}_{xy}\right)
\left(1; (x-y)^\mu (x-y)^\nu\right)\per
\label{eqsunset}
\eea
The integrals proportional to $\triangle^3_{xy}$ correspond to
the sunset integrals $H$ and $H^{\mu\nu}$ discussed in appendix
\ref{apdiagrams}. The integrals that contain
$\dot{\triangle}_{xy}$ can be evaluated in a similar manner.
Proceeding in an analogous fashion for
the remaining contributions to $Z_2$, we finally find that the
following
counterterms render the two--loop functional finite at $d=4$:
\bea
b_1\!\!\!&=&\!\!\! \left(\frac{\mu^{2w}}{16 \pi^2}\right)^2
\left\{ \frac{(2+N)}{2w}+b_1^r(\mu,w)\right\}\co\nn
b_2\!\!\!&=&\!\!\! \left(\frac{\mu^{2w}}{16 \pi^2}\right)^2
 \left\{
\frac{(5+N)(2+N)}{w^2}
+\frac{(2+N)}{w}\left(3\!-\!a_2^r(\mu,w)\!-\!a_3^r(\mu,w)
\right) \!+\! b_2^r(\mu,w)  \right\}
\co\nn
 b_3\!\!\!&=&\!\!\! \left(\frac{\mu^{2w}}{16 \pi^2}\right)^2
\left\{
\frac{(8+N)^2}{w^2} +
\frac{2}{w}\left(22+5N-(8+N)a_3^r(\mu,w)\right)+b_3^r(\mu,w)
\right\}  \label{eqcount2}\per \nn\eea
{\underline{Remark:}} In Eq.~(\ref{eqcount2}), the
full contribution $a_{2,3}^r(\mu,w)$ to the singular part in
$b_{2,3}$ occurs. We could have expanded $a_{2,3}^r(\mu,w)$
around $w=0$
 in (\ref{eqa_i}) -- then, only $a_{2,3}^r(\mu,0)$ would occur
 in (\ref{eqcount2}). In order to keep $a_{2,3}$ scale
independent, additional terms of order $O(w)$ must then be
added in (\ref{eqa_i}). From
 (\ref{eqcount2})  it is clear that these  would
contribute  a local part to $Z_2$ that could then be removed
with
a redefinition of the renormalized couplings $b_i^r(\mu,w)$.
The
definition (\ref{eqa_i})
allows one to do all this  in one go.

\subsubsection{The role of the equations of motion}
We come back to the choice of the counterterms $C_i$\footnote{
The following discussion is adapted to
CHPT, where
the higher--order lagrangians ${\cal L}_4,{\cal L}_6,\ldots$ take
the role of the counterterms $C_1,C_2,\ldots.$}.
In  the generating functional, they are evaluated at the solution
to the EOM. This amounts to a linear relation among the
$\overline{P}_i$,
\bea
E&\doteq&
2{P}_1+2{P}_2+4{P}_3
+{P}_4\scs\nn
 \overline{E}&=&0\per\label{eqeompi}
\eea
As we already mentioned, one may therefore remove one of the
$P_i$ from the list  (\ref{eqp_i}). We may e.g., eliminate
$P_3$ instead of  $P_4$ as above,
\bea
C_1\rightarrow C_1'=ga'^TP'\scs P'^T=(P_1,P_2,P_4)\per\label{eqnewc1}
\eea
 Using the identity
\bea
a'^TP'&=&c^TP+a_3' E\scs\nn
c^T&=&(a_1'-2a_3',a_2'-2a_3',-4a_3')\scs\label{eqident}
\eea
it is seen that a different choice of the counterterm at
one--loop order amounts to adding to the
lagrangian a term that vanishes at the solution to the EOM.
As we now demonstrate, the use of $C_1'$
 requires that the counterterm at two--loop order must be
adapted accordingly,
\bea
C_2&\rightarrow& C_2'=g^2b'^TP\per\label{eqnewc}
\eea
In $C_2'$,
we  use again $P$ -- the freedom in the choice of
polynomials occurs at each order in the perturbative expansion.
We denote by $Z_1',Z_2'$ the generating
functional obtained with  (\ref{eqnewc1}, \ref{eqnewc}). For
the one--loop functional  to be
finite at $d=4$, the integrals
$\int dx\, \overline{C}'_1$ and $\int
dx\, \overline{C}_1$ must  have the same singularities.
 According to the relation (\ref{eqident}), this requirement is
satisfied with
\bea
{a'}^T=(-\frac{a_3}{2},a_2-\frac{a_3}{2},-\frac{a_3}{4
} )\per \label{eqaap1}
\eea
The replacement $C_1\rightarrow C'_1$
 then amounts to
\bea
C_1\rightarrow C_1+\kappa g E\scs\label{eqtrans}
\eea
with $\kappa=-a_3/4$. This transformation
obviously leaves the one--loop functional
 unchanged. Repeating the calculation of $Z_2'$ with
(\ref{eqtrans}), one finds
\bea
Z_2'=Z_2 +\int dx\, \left(\ol{C}_2'-\ol{C}_2\right)
+\overline{Q}\scs \pnn
 \eea
where $Q$ is a local action,
\bea
Q&=&-g^2\kappa\int dx e^TP\scs\nn
e^T&=&(2a_1+\kappa,2a_2+\kappa,4a_3+6\kappa)\per
\eea
In addition to the singularities in $Z_2$, the counterterm
$C_2'$ has
to cancel  the ones in $\overline{Q}$ as well. Because
the latter is a local
action, it can  be removed from $Z_2'$ altogether
with
 \bea
b'= b+\kappa e\per\pnn
\eea
In summary, a different choice of the counterterm at one--loop
order amounts
to adding a term to the lagrangian that vanishes at the solution
to the EOM, see Eq. (\ref{eqtrans}). That
transformation  leaves $Z_1$ untouched and changes
$Z_2$ by a local
action $\ol{Q}$ which can be completely removed by an appropriate
choice of the counterterm at two--loop order,
\bea
\left(
\begin{tabular}{l}
$C_1\rightarrow C_1+\kappa g E$  \\
$C_2\rightarrow C_2+g^2e^TP$ \end{tabular}\right)
 \rightarrow
Z_1'=Z_1, Z_2'=Z_2\per
\eea
\subsubsection{Connection with the MS scheme}
The above renormalization scheme generates scale--independent
quantities at each step. In particular, individual Feynman
diagrams are independent of $\mu$. The Green functions so
evaluated at one--loop accuracy contain the finite parameters
$M^2,g, a_i(\mu,0),b_i(\mu,0),\ldots \,$\footnote{Of course,
since the present theory is renormalizable, these parameters
occur in such a combination that they can  be lumped into,
e.g., the physical mass $M_P$
and the physical coupling constant.}.  An analogous procedure is
used
in chiral perturbation theory \cite{glann}. We find it useful
to compare at this stage this scheme with the MS scheme
 in conventional perturbative calculations.  There, one starts
{}from the bare lagrangian
\bea\label{eqlbare}
{\cal L}=
\frac{1}{2}(\partial_\mu \phi^T_B\partial^\mu\phi_B
-M_B^2\phi^T_B\phi_B) -\frac{g_B}{4}(\phi^T_B\phi_B)^2
\eea
and sets at one--loop order
\bea\label{eqgbare}
g_B=\mu^{-2w}g_r(\mu)\left\{1-
\frac { ( N + 8 ) } {16\pi^2w} g_r(\mu) +O(g_r^2)\right\}\co\nn
M^2_B=M_r^2(\mu)\left\{1-\frac{(N
+ 2
) } { 1 6 \pi^2w}g_r(\mu) +O(g_r^2)\right\}\per\label{eqms1}
 \eea
The two schemes can
 easily be related by evaluating, e.g., the physical mass $M_P$
and the
elastic scattering amplitude  to
one--loop accuracy. This is done in the following subsection in
the present scheme. Repeating that calculation
with (\ref{eqlbare}, \ref{eqgbare})
and  identifying the one--loop expressions gives
\bea
g_r(\mu)=g(1+\frac{ga_3^r}{16\pi^2} +O(g^2))\scs\nn
M^2_r(\mu)=M^2(1+\frac{ga_2^r}{16\pi^2}+O(g^2))\per\pnn
\eea
Here and in the following we use the notation
\bea
a_i^r\doteq a_i^r(\mu,0) \scs b_i^r\doteq b_i^r(\mu,0)\per\pnn
\eea
{}From the scale dependence of the renormalized couplings
$a_2^r,a_3^r$ one reads off the scale dependence of $M_r^2,g_r$
that guarantees scale independence of physical quantities.
Analogous relations hold to all orders in the perturbative
expansion.

\subsection{Two-- and four--point functions}
For illustration, we evaluate the two--and four--point functions.
It allows us to explain the methods used in Ref.
 \cite{BCEGS1} for the evaluation of the $\pi\pi$ scattering
amplitude in CHPT.

\subsubsection{Two--point function}
\begin{figure}[t]
\begin{center}
\mbox{\epsfysize=10cm \epsfbox{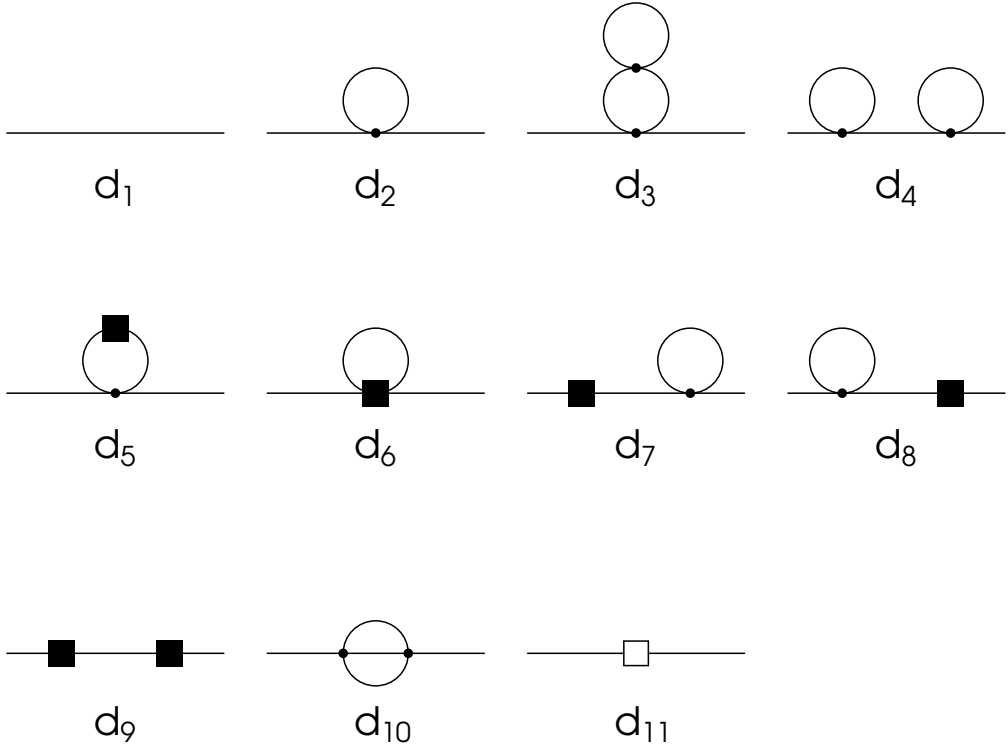}
     }
\caption{Diagrams that contribute to the two--point function
at two--loop order. A filled
square (open square) stands for the contributions from the
counterterm $C_1 (C_2).$ } \label{f2} \end{center}
\end{figure}

To evaluate the two--point function
\bea
i\langle
0|T\phi^a(x)\phi^b(y)|0\rangle_{conn}&=&
\delta^{ab}\triangle'_{xy}\nn
&=&\delta^{ab}\left(N^{-1}\langle\triangle_{xy}\rangle+O(g)
\right ) \scs \eea
 we may either calculate the relevant
Feynman diagrams displayed in Fig. \ref{f2}, or determine
the contribution of order $f^2$ to the generating functional,
because
\bea      \label{eqprop}
\delta^{ab}\triangle'_{xy}=\left.\frac{\delta^2 Z}{\delta
f^a_x\delta f^b_y}\right|_{f=0}\per
\eea

We first proceed in this manner and find
 \bea
Z&=&\frac{1}{2}\int dx\, \Pc^T_xK_{xy}\Pc_y +
O(\hbar^3,\Pc^4)\scs\nn
K_{xy}&=&\delta^4(x-y)(A\Box_y+B)
+C\triangle_{xy}+D\triangle_{xy}^3\nonumber
\eea
with
\bea
A&=&1-g^2 b_1\scs\nn
B&=& M^2-g\, m^2
  - g^2\left\{ b_2 M^2 + (N+2) ( a_3 T_M + m^2
  \dot{T}_M ) \right\} \scs\nn
C&=& g^2\, m^4\scs
D=-2 g^2 (N+2)\scs\nn
m^2&=&a_2 M^2+(N+2)T_M  \per
\eea
Next we  consider the Fourier transform
 \bea
\triangle'(p^2)&=&\int d^dx e^{-ip(x-y)}\triangle'_{xy}\nn
&=&\frac{1-g^2b_1}{M^2-p^2-B+M^2A
-2g^2(N+2)H(p^2)}+O(g^3)\scs\label{eqmp}
\eea
where $H(p^2)$ denotes the sunset integral  (\ref{eqa14}).
Expanding $H(p^2)$ around $p^2=M^2$ gives at $d=4$
\bea
{\triangle}'(p^2)=\frac{Z_{\phi}}{M_P^2-p^2}+R_\phi(p^2)+O(g^3)\scs
\eea
with
\bea
Z_{\phi}&=&1+g^2
\frac{1}{(16\pi^2)^2}\left( (N+2)( \frac{3}{4}+a(M/\mu))
-b_1^r\right) +O(g^3)\scs
\eea
and
\bea
R_\phi(p^2)=2g^2(N+2)\ol{\ol{H}}(p^2)\per
\eea
An integral representation for the twice subtracted
sunset integral  $\ol{\ol{H}}(p^2)$ -- which is regular at
$p^2=M_P^2$ -- is given
in (\ref{eqa50}). Finally, the
physical mass $M_P$
can be easily evaluated from Eqs. (\ref{eqmp}, \ref{eqa51}) to
two loops. The  expansion starts with
\bea\label{eqmassg2}
M_P^2&=&M^2+gm^2+O(g^2)\nn
&=&M^2\left\{1+\frac{g}{16\pi^2}[(N+2)(a(M/\mu)-1)+a_2^r]+O(g^2)
\right\}\per \eea

\begin{figure}[t]
\begin{center}
\mbox{\epsfxsize=10cm \epsfbox{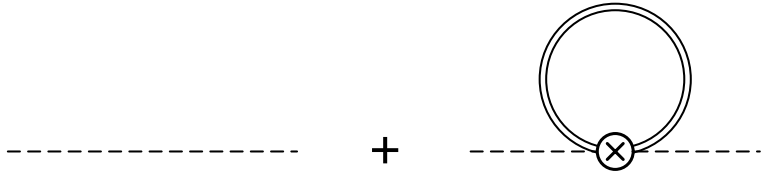}
     }
\caption{Summing up the graphs d$_1\cdots$ d$_9$ in Fig.
\protect\ref{f2}. See text after Eq. (\protect\ref{eqmassg2}).}
 \label{f3} \end{center}
\end{figure}

We add a remark concerning the calculation of the two--point
function from the Feynman diagrams displayed in Fig.
 \ref{f2}. Instead of evaluating the diagrams
d$_1,\ldots,$d$_9$ individually, one may
calculate the tree and tadpole graphs shown in Fig. \ref{f3},
where the dashed and double lines stand for the propagators
\bea
\frac{1}{M_1^2-p^2}\scs M_1^2&=&M^2[1+ga_2+g^2(b_2-b_1)]
\hspace{.5cm}
\mbox{(dashed line)}\nn
\frac{1}{M_2^2-p^2}\scs M_2^2&=&M^2+gm^2
\hspace{.5cm}
\mbox{(double--line)}\nn
\nonumber
\eea
and where the vertex denotes the coupling $g(1+a_3g)$. The sum of
these two diagrams is
\bea\label{eqmpi}
\frac{1}{M_3^2-p^2}\scs&& M_3^2=M_1^2+g(1+a_3g)(N+2)T_{M_2}\scs
\eea
where the argument in the tadpole integral  is $M_2$.
The expression (\ref{eqmpi}) is equal to the sum of the graphs
d$_1,\ldots,$d$_9$ up to terms beyond two loops.

\subsubsection{Four--point function}
\begin{figure}[t]
\begin{center}
\mbox{\epsfysize=12cm \epsfbox{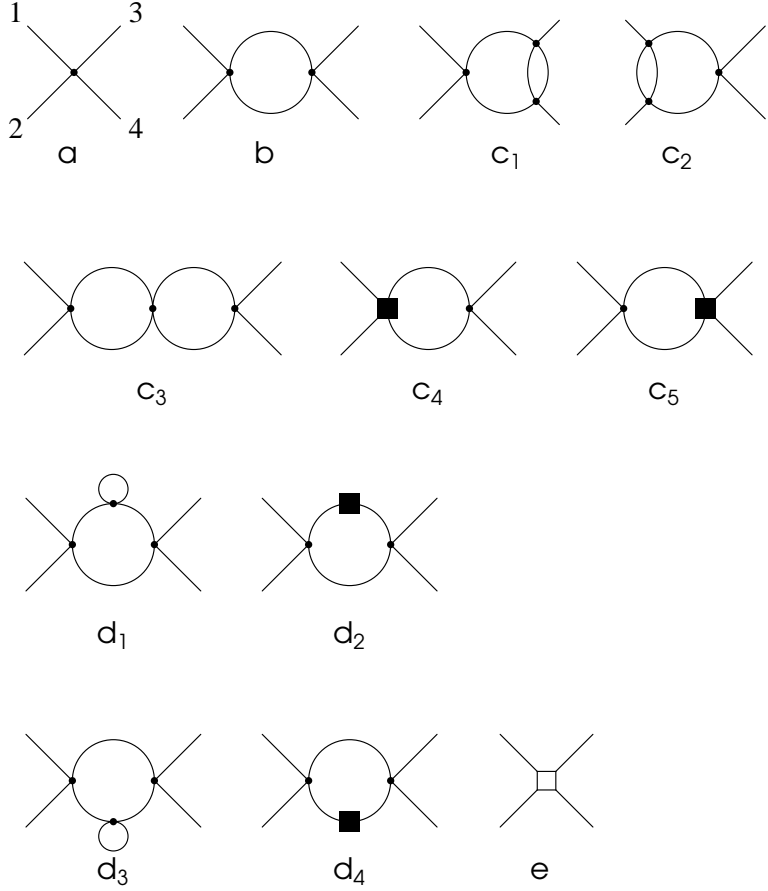}
     }
\caption{The elastic $\phi\phi\rightarrow\phi\phi$ scattering
amplitude to two loops. The numbers on the external lines denote
the momenta. Crossed diagrams and insertions on external lines
are not displayed. A filled box (open box) denotes contributions
{}from the counterterm $C_1 (C_2)$.} \label{f4} \end{center}
\end{figure}

The number of diagrams at two--loop order becomes quite large --
of course even more so in the case of CHPT. In the case of the
two--point function, we have just seen that almost all graphs in
Fig. \ref{f2} can be summed up by evaluating the
tadpole diagram shown in Fig. \ref{f3}. Here we
wish to illustrate an analogous method to evaluate
the elastic scattering amplitude (four--point
function) to two loops. Whereas the method does not really pay
 off in the case of the
$N$--component $\phi^4$ theory
considered here, it turns out to be very useful in  CHPT. Furthermore,
a similar procedure works in the
evaluation of form factors\footnote{This method has already been used
in \cite{BGS94,Bur96} in connection with the evaluation of the
process $\gamma\gamma\rightarrow\pi\pi$ to two loops.}.
The four--point function is of the form
\bea
&&i^3\int dx_1\,dx_2\,dx_3\,
e^{-i(p_1x_1+p_2x_2-p_3x_3-p_4x_4)}\langle0|T\phi_i(x_1)
\phi_k(x_2)\phi_l(x_3)\phi_m(x_4)|0\rangle\nn
&&=\frac{Z_{\phi}^2}{\prod_i(M_P^2-p_i^2)}T_{lm;ik}(s,t,u;p_1^2,p
_2^2,p_3^2,p_4^2)\scs\label{eq4p}
\eea
with
\bea
&&p_1+p_2=p_3+p_4\co\nn
&&s=(p_1+p_2)^2\scs t=(p_1-p_3)^2\scs u=(p_1-p_4)^2\per\pnn
\eea
The scattering amplitude is obtained by putting all momenta on
the mass shell,
\bea
T_{lm;ik}(s,t,u;p_1^2,p_2^2,p_3^2,p_4^2)|_{p_i^2=M_P^2}
=
\delta_{ik}\delta_{lm}A(s,t,u) +\mbox{cycl.}
\per\nonumber
\eea
To determine the amplitude $A(s,t,u)$, it suffices to calculate
the matrix element for the indices
\bea
i=k=1\scs m=l=N\per
\eea
The relevant graphs up to $O(g^3)$ are displayed in Fig.
 \ref{f4}.
The numbers attached on the external lines denote the relevant
momenta, $i\leftrightarrow p_i$. The group indices and crossed
diagrams are not shown, and  mass and counterterm insertions on
the external lines are not displayed.

\begin{figure}[t]
\begin{center}
\mbox{\epsfxsize=12cm \epsfbox{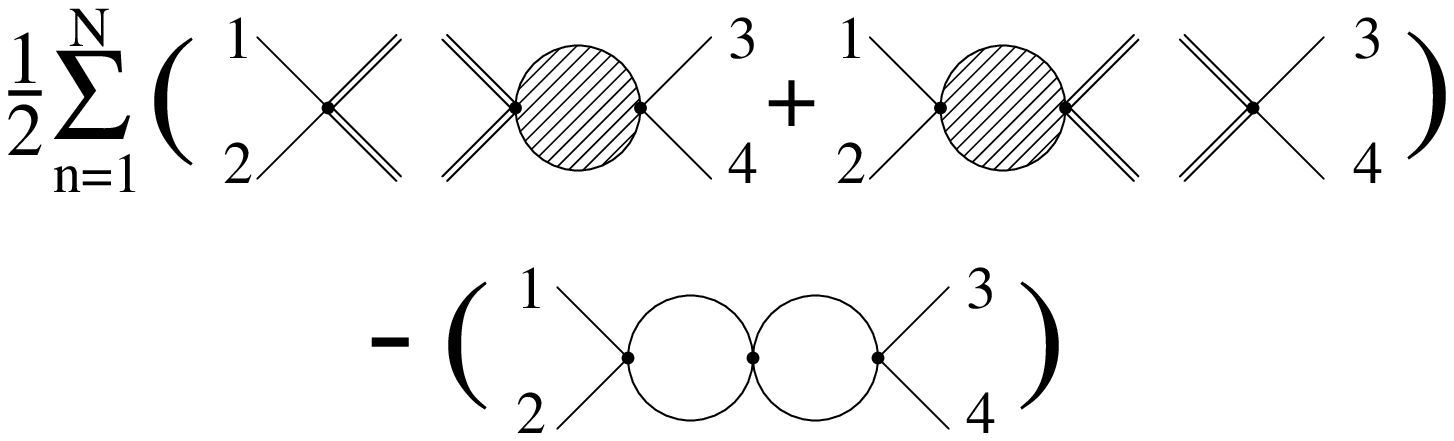}
     }
\caption{Summing up the diagrams c$_1\cdots$ c$_5$ in Fig.
\protect\ref{f4}. See text after Eq.
(\protect\ref{eqscatta}).} \label{f5} \end{center}
\end{figure}

We use the obvious notation
\bea
A(s,t,u)=gA^{(1)}+g^2A^{(2)}+g^3A^{(3)}+O(g^4)\per
\eea
The tree--level and one--loop results are
\bea\label{eqscatta}
A^{(1)}(s,t,u)&=&-2\scs\nn
A^{(2)}(s,t,u)&=&  -\frac{2}{16\pi^2}
  \left[a_3^r+(N+8) a(M/\mu) \right]\nn
& +& 2(N+4)\bar{J}(s)+4
  \left[\bar{J}(t)+\bar{J}(u) \right] \co \nn
\eea
with
\bea
 \bar{J}(z){=}\frac{1}{16\pi^2}
\left[\sigma \ln \frac{\sigma-1}{\sigma+1}+2\right]
\;\;\; ; \; \;\;
\sigma=\left(1-4M^2/{z}\right)^\frac{1}{2}\per
\eea
 The calculation at two--loop order may be made
more economical by using the renormalized one--loop off--shell
amplitude as a single (nonlocal) vertex. Then, a large part of
the two--loop diagrams can be obtained automatically by making
a one--loop calculation where one of the vertices is the
one--loop amplitude, and the other is the standard tree--level
vertex. To be more specific, consider the diagrams  Fig.
 \ref{f4}c$_1$,\ldots,c$_5$. They can be calculated by
evaluating the integrals indicated in Fig. \ref{f5}.
The notation is as follows: The double lines denote
off--shell legs (the corresponding off--shell amplitude is
defined in (\ref{eq4p})). These legs carry momenta $l$ and
$(p_1+p_2-l)$, and group index $n$. The integration has to be
done with weight
\bea
\frac{1}{(M^2-l^2)(M^2-(p_1+p_2-l)^2)}\per\pnn
\eea
{}From this sum, one has to subtract the diagram Fig.
 \ref{f4}c$_3$. The result equals the diagrams Fig.
 \ref{f4}c$_1$,\ldots,c$_5$.

One may easily check the correctness of this statement, if
one singles out of the full one--loop amplitude one
particular diagram. In this manner one generates a single
two--loop diagram with a symmetry factor in front of it: One
can then compare this to what one would get from a standard
Feynman diagram calculation. Similar formulae hold, of course,
for the crossed versions of these diagrams.

After checking that the method is correct, the second
question to be answered is how one can do Feynman loop
integrals with a nonlocal vertex in it. In particular, how
does one perform the following integral:
\bea
 \int
\frac{d^dl}{(2\pi)^d}\frac{\bar{J}((l-p_3)^2)}{(M^2-l^2)(M^2
-(p_1+p_2-l)^2)}\label{eqnfv}\per
\eea
First we remark that this integral has to be calculated
anyway, even if one would try to calculate directly the
two--loop diagrams. It corresponds to the fish diagram in
Fig. \ref{f4}c$_1$, where the divergence of the one--loop
subdiagram has already been cancelled by the insertion of
the counterterm Fig. \ref{f4}c$_5$. As can be seen here, one of the
advantages of the method is that one starts the calculation
at a stage where the one--loop subdivergence has already
been subtracted.\footnote{The subtraction of the
subdivergence in the diagram Fig. \ref{f4}c$_3$ is much
easier to accomplish, because the corresponding integral
is a product of two one--loop integrals.} This fact is of
relevance in particular in CHPT, where one may thus first do
the renormalization of the off--shell one-loop amplitude
before diving into the forest of two--loop integrals. Next
we note that the loop function $\bar{J}(s)$ admits a
 dispersive representation in $d$ dimensions
\cite{BGS94,Bur96}:
 \bea
\bar{J}(t)=t
\int_{4M^2}^{\infty}\frac{[d x ]}
{ x ( x -t)}\scs
\eea
where the measure $[d x ]$ is given in appendix
\ref{apdiagrams} (for $M^2=1$). By inserting this expression into
Eq. (\ref{eqnfv})
we see that the problem is transformed into the calculation of
a one--loop integral with three propagators, where one of
the propagators has a variable mass over which we finally
have to integrate:
\bea
\int_{4M^2}^{\infty}\frac{[d x ]}{ x }\int
\frac{d^dl} {
( 2 \pi)^d}\frac{(l-p_3)^2}{(M^2-l^2)(M^2-(l-p_3-p_4)^2)( x -
(l-p_3)^2)}\per
\eea
We discuss in  appendix \ref{apdiagrams} how this integral can be
done explicitly \cite{BP68}.

Finally, we are left with the graphs that we have not yet
taken into account:  The tadpole and counterterm insertions
displayed in Fig. \ref{f4}d$_{1,\ldots,4}$
can easily be taken care of by simply replacing in the
one--loop integrals the parameter $M$ with the physical mass
$M_P$.
 The contributions
{}from  wave function renormalization is taken care of by
multiplying the tree graph with $Z_{\phi}^2$.

In summary, the advantages of the approach just described
are that i) by using the renormalized one--loop amplitude as
a single nonlocal vertex, the divergences due to one--loop
subgraphs in nonfactorizable integrals never appear; ii) the
diagrams that produce the mass renormalization inside the
one--loop amplitude are taken into account trivially and do
not have to be calculated explicitly; iii) the method also
applies to the calculation of form factors, or any process
where one part is given by an elastic two--body process like
$\gamma\gamma\rightarrow \pi\pi$ \cite{BGS94}; iv) the method can
easily be implemented into a computer
algebra program like FORM \cite{FORM}.

\setcounter{equation}{0}
\section{CHPT to two loops}

In the following we discuss the methods used in Ref.
\cite{BCEGS1}
to evaluate the elastic $\pi\pi$ scattering amplitude to two
loops
in the framework of CHPT. This section contains details
concerning renormalization, whereas the two--loop expressions for
the pion mass, the pion decay constant and the scattering
amplitude are given in the following section.

\subsection{The lagrangian}
The effective lagrangian consists of a string of terms,
\bea
         {\cal L}_{\mbox{\tiny{eff}}} = {\cal L}_2 + \hbar {\cal
         L}_4 + \hbar^2 {\cal L}_6 + \dots .
\label{eqstring}
\eea
Depending on the application one has in mind, one has to keep
external fields in ${\cal L}_{\mbox{\tiny{eff}}}$.
 In the present case,
we wish to calculate $M_\pi,F_\pi$ and the $\pi\pi$ scattering
amplitude. The pion mass and the scattering amplitude may both be
calculated from on--shell quantities, and external fields are
thus not necessary. On the other hand, the pion decay constant
needs a weak current as an external probe -- it is an off--shell
quantity, not accessible through on--shell matrix elements. The
use of the axial current to evaluate $F_\pi$ to two loops is
cumbersome because of the presence of Lorentz indices.
 In the following, we use instead the two--point function
of pseudoscalar densities. A Ward identity relates the residue of
this quantity to  the pion decay constant.

There are many choices for the pion fields to be used in the
effective lagrangians. Of course, the result for physical
quantities is always the same. A convenient choice to minimize
the number of diagrams is the sigma model parametrization. In the
following we work in the isospin symmetry limit $m_u=m_d$. We
have
           \bea
         {\cal L}_2 = \frac{F^2}{4} \left\la u_\mu u^\mu+
 \chi_+\right\ra\co
         \eea
with
         \bea\label{eqsigma}
         U &=& \sigma + i \frac{\mbox{\boldmath$\phi$}}{F} \scs
\sigma^2 + \frac{\mbox{\boldmath$\phi$}^2}{F^2} = \unith \co
         \nn
        \mbox{\boldmath$ \phi$} &=&
 \left( \mbox{$\begin{array}{cc} \pi^0
& \sqrt{2} \pi^+
         \\ \sqrt{2} \pi^- &- \pi^0 \end{array}$} \right) =
\phi^i \tau^i  \co\nn
         u_\mu &=& i u^\dagger \partial_\mu Uu^\dagger = -i u
\partial_\mu U^\dagger u = u_\mu^\dagger \co
         \nn
         \chi_+ &=& u^\dagger \chi u^\dagger + u \chi^\dagger u
\co \nn
         \chi &=& 2 B(\hat{m}\unith+ i p)\scs
\hat{m}=\frac{1}{2}(m_u+m_d)\co
 \eea
with $u^2=U$. Furthermore, $p=p^i\tau^i$ is the external
pseudoscalar field referred to above. The symbol $\la A \ra$
denotes the trace of the two--by--two matrix $A$.
The lagrangian ${\cal L}_4$ is \cite{glann}
  \bea
         {\cal L}_4 &=& \sum^{4}_{i=1} l_iP_i + \dots \co
         \label{eqleff4}
         \eea
         where
         \bea
         P_1 &=& \frac{1}{4} \la u^\mu u_\mu \ra^2 \scs
         P_2 = \frac{1}{4} \la u_\mu
         u_\nu\ra \la u^\mu u^\nu \ra \scs
         P_3 = \frac{1}{16} \la \chi_+\ra^2 \scs
         P_4 = \frac{i}{4} \la u_\mu
         \chi^\mu _- \ra \scs\nn
         \eea
with
         \bea
         \chi_-^\mu&=& u^\dagger \partial^\mu\chi u^\dagger
-u\partial^\mu\chi^\dagger u \per
\eea
The ellipsis in Eq. (\ref{eqleff4}) denotes terms that do not
contribute to the quantities considered here.  The low--energy
constants $l_i$ are divergent and remove the ultraviolet
divergences generated by one--loop graphs from ${\cal L}_2$. We
discuss them in more detail below.

The complete effective lagrangian ${\cal L}_6$ is not yet
available: Whereas the list of the necessary counterterms has
been published \cite{scherer}, their divergence structure at
$d=4$
is still under investigation \cite{Betal97}. The knowledge of these
would provide
us with a check on the calculation, because these
divergences
must cancel the ones that we find in the two--loop
calculation. Apart from this check, however, that analysis would
not be of further use in the present context, because the
scale--dependent finite pieces of those couplings are still largely
unknown. Nevertheless, the structure displayed in \cite{scherer}
shows that there are no algebraic constraints  between
the counterterms
at $O(p^6)$ needed in the expressions for the pion mass, pion
decay constant or $\pi\pi$ scattering amplitude.

For the following considerations, it is useful to recall that the
expansion in powers of the momenta is equivalent to an expansion
in inverse powers of $F^2$. In an obvious notation, the chiral
expansion for the pion mass, the pion decay constant and the
elastic $\pi\pi$ scattering amplitude is
\bea
M_\pi^2&=&M^2\left\{1+m_4 \frac{M^2 }{F^2} +m_6\frac{M^4}{F^4}
+O(F^{-6})\right\}\co\nn
F_\pi&=&F\left\{1+f_4 \frac{M^2 }{F^2} +f_6\frac{M^4}{F^4}
+O(F^{-6})\right\}\co\nn
A(s,t,u)&=&\frac{A_2}{F^2} +\frac{A_4}{F^4}
+\frac{A_6}{F^6} +O(F^{-8})\per\label{eqexpand}
\eea
The one--loop contributions $m_4,f_4$ and $A_4$ have been
determined  in \cite{glpl} -- here, we calculate
 $m_6,f_6$ and $A_6$. Therefore, we need to keep in the
effective lagrangian only contributions up to $O(F^{-6})$. We
write symbolically
\newcommand{\lag}[2]{ \frac{\mbox{\boldmath$\phi$}^{#1}}{F^{#2}}}
\bea
{\cal L}_2&=&{\cal L}_{\small{kin}}+a_1\lag{4}{2}+a_2\lag{6}{4}
+a_3\lag{8}{6} +O(F^{-8})\co\nn
{\cal L}_4&=&\hspace{1.34cm}b_1\lag{2}{2}+b_2\lag{4}{4}
+b_3\lag{6}{6} +O(F^{-8})\co\nn
{\cal L}_6&=&\hspace{2.8cm}c_2\lag{2}{4}+c_3\lag{4}{6}+O(F^{-8})
\co
\eea
where the dimension of the couplings  is
\bea
[a_i]=\mbox{mass}^2\scs [b_i]=\mbox{mass}^4\scs [c_i]=\mbox{mass}^6\per
\eea
In order to generate a term of order $F^{-2m}$, one has to
consider all products
\bea
a_{n_1}^{m_1}b_{n_2}^{m_2}c_{n_3}^{m_3}\;\;\; ; \;\;\;
m_1n_1+m_2n_2+m_3n_3=m\per
\eea
Furthermore, it is very convenient to collect the terms
quadratic in the pion fields in the kinetic part and to expand
in powers of $b_1$ and $c_2$ only afterwards. As a result, one
has to consider the following products:
\bea
\begin{array}{rllllll}
m_6,f_6:&a_1^2;&a_2;&b_2\\
A_6:&a_1^3;&a_1a_2;&a_3;&a_1b_2;&b_3;&c_3\;.
\end{array}\label{eqcount}
\eea

\subsection{Renormalization at two--loop order}
\subsubsection{General analysis}
The loop contributions to $M_\pi,F_\pi$ and to the scattering
amplitude $A(s,t,u)$ are divergent in the limit $d\rightarrow 4$.
Using  the notation (\ref{eqexpand}), we are concerned here
with the renormalization of the quantities $m_6,f_6$ and
$A_6$. As is guaranteed by general theorems of renormalization
theory \cite{Collins}, the divergent parts of $m_6$ and $f_6$ are mass
independent in dimensional regularization. Likewise, the
divergent part of $A_6$ is a polynomial in the external momenta
and in $M^2$. The most general crossing symmetric polynomial
arising at order $p^6$ is
\bea
F^6A_6^{\mput{pol}}&=&
A^{(1)}M^6+A^{(2)}sM^4+A^{(3)}s^2M^2+
A^{(4)}(t-u)^2M^2\nn
&&+A^{(5)}s^3+A^{(6)}s(t-u)^2\pnn
\eea
with six coefficients $A^{(1)},\ldots,A^{(6)}.$ Like $m_6$ and
$f_6$, they receive contributions from two--loop graphs with
${\cal L}_2$, one--loop graphs with one vertex from ${\cal L}_4$
and tree graphs generated by ${\cal L}_2+{\cal L}_4+{\cal L}_6$,
see Eq. (\ref{eqcount}).
The following analysis applies to each of the eight coefficients
$m_6,f_6,A^{(1)},\ldots,A^{(6)}$ separately. Denoting these
coefficients generically as Q, we write
\bea
Q=Q_{\mput{loop}}+Q_{\mput{tree}}\per\label{eqq}
\eea
Since $Q$ is part of a measurable quantity, it must of course be
finite (and scale independent). The two components
$Q_{\mput{loop}}$ and $Q_{\mput{tree}}$, on the
other hand, are both divergent. Concentrating first on
$Q_{\mput{loop}}$, dimensional
regularization produces for $m_6$ and $f_6$ the general
form
\bea
Q_{\mput{loop}}&=&\dot{T}_M\left\{\dot{T}_Mx(d)-\sum_{i=1}^
{4}l_i(d)y_i(d)\right\}\co\label{eqqloop}
\eea
where $x(d),y_i(d)$ are dimensionless functions of $d$,
finite at $d=4$. (For the sake of clarity, we  exhibit
in this section the dependence of the low--energy couplings
$l_i(d)$
on the space--time dimension $d$.) The same structure is found
for the six
coefficients $A^{(i)}$. We  perform a Taylor series expansion,
\bea
x(d)&=&x_0+x_1w+x_2w^2+O(w^3) \co\nn
y_i(d)&=&y_{i0}+y_{i1}w+y_{i2}w^2+O(w^3)\;\;\; ; i=1,\ldots
,4\;\; ; \;\;w=\frac{d}{2}-2\pnn
\eea
with real numbers $x_0,x_1,x_2,y_{i0},y_{i1},y_{i2}$.
The tadpole integral $\dot{T}_M$ (\ref{eqtaddot}) and the low--energy
constants $l_i(d)$ are both divergent as $w\rightarrow 0$. The
Laurent expansion of the tadpole integral is displayed in (\ref{eqtadp}).
For the low--energy constants $l_i(d)$, we write
\bea
l_i(d)&=&\frac{\mu^{2w}}{(4\pi)^2}\left[\frac{\gamma_i}{2w}+
l_{i,r}^{\mput{MS}}(\mu,w)\right]\co\nn
\gamma_1&=&\frac{1}{3}\scs \gamma_2=\frac{2}{3}\scs
\gamma_3=-\frac{1}{2}\scs \gamma_4=2\per\label{eqexpl}
\eea
For reasons already explained in Sect. 2 in the framework of the
$N$--component $\phi^4$ theory,
we have not expanded the renormalized couplings
$l_{i,r}^{\mput{MS}}(\mu,w)$ around $w=0$. See comments after
Eq. (\ref{eqcount2}) and the discussion below.

Since there must not be any terms of the form
\bea
\frac{\ln M/\mu}{w}\pnn
\eea
in $Q_{\mput{loop}}$, there are eight consistency conditions
(for each of the eight coefficients $m_6,f_6,A^{(i)}$) of the
type \cite{wein79}
\bea
x_0=\frac{1}{4}\sum_{i=1}^{4}\gamma_iy_{i0}\per\label{eqlog}
\eea
This absence of mass singularities in the divergences provides an
extra check on our calculation. Using the equality
(\ref{eqlog}) that relates one-- and two--loop parameters, one
gets (from now on, the summation convention is implied for
$i=1,\ldots,4$)
\bea
Q_{\mput{loop}}&=&\frac{\mu^{4w}}{(4\pi)^4}\left\{\frac{Q_2}{w^2}
+\frac{Q_1}{w}+Q_0+O(w)\right\}\co\nn
Q_2&=&-x_0\scs Q_1=
x_1-l_{i,r}^{\mput{MS}}(\mu,w)y_{i0}-
\frac{1}{2}\gamma_iy_{i1}\co\nn
Q_0&=&
 x_0a(M/\mu)^2+\left[2x_1-l_{i,r}^{\mput{MS}}(\mu)y_{i0}
-\frac{1}{2}\gamma_iy_{i1}\right]a(M/\mu)\nn
&&+x_2-l_{i,r}^{\mput{MS}}(\mu)y_{i1}-\frac{1}{2}\gamma_iy_{i2}
\co\pnn
\eea
where
\bea
l_{i,r}^{\mput{MS}}(\mu)\doteq l_{i,r}^{\mput{MS}}(\mu,0)\per\pnn
\eea
It is seen that the coefficient $c(M/\mu)$ in the Laurent
expansion (\ref{eqtadp}) of the tadpole integral $\dot{T}_M$
does not enter due to the relation (\ref{eqlog}).

The pole terms can now be absorbed by the tree--level
contribution $Q_{\mput{tree}}$ which is a certain combination
(depending on the specific observable under consideration) of
coupling constants in the general chiral lagrangian ${\cal L}_6$.
 Denoting this combination generically as $z(d)$, the
expansion analogous to (\ref{eqexpl})  is
\bea
Q_{\mput{tree}}(d)\doteq z(d)=
\frac{\mu^{4w}}{(4\pi)^4}\left\{-\frac{Q_2}{w^2}-\frac{Q_1}{w}
+(4\pi)^4
z_r^{\mput{MS}}(\mu)+O(w
)\right\}\per\label{eqzd}
\eea
The sum $Q_{\mput{loop}}+Q_{\mput{tree}}$ is finite and
independent of the scale $\mu$ by construction. Taking the limit
$d\rightarrow 4$ yields
\bea
\lim_{d\rightarrow 4}
Q=\frac{Q_0}{(4\pi)^4}+z_r^{\mput{MS}}(\mu)\per\label{eqqms}
\eea
Recalling the scale independence of $Q,\dot{T}_M, l_i(d) $ and
 of $z(d)$, one derives the renormalization
group
equation for the renormalized coupling constant
$z_r^{\mput{MS}}$:
 \bea
\mu\frac{dz_r^{\mput{MS}}(\mu)}{d\mu}=
\frac{2}{(4\pi)^4}\left[2x_1-l_{i,r}^{
\mput{MS}}(\mu)y_{i0}-\gamma_iy_{i1}\right]\per\label{eqzmrs}
\eea
Comparing Eqs. (\ref{eqtada}), (\ref{eqlog}) and (\ref{eqqms}),
we
recover the well--known fact \cite{wein79, BGS94, colangelo} that
the coefficient of the
double logs can be determined solely from one--loop diagrams with
a single vertex from ${\cal L}_4$.
\subsubsection{Modified minimal subtraction}
The above renormalization procedure  corresponds to
minimal subtraction, where powers of $\ln 4\pi+\Gamma'(1)$ occur
in the final expressions, see (\ref{eqtada}). These terms can be
absorbed by splitting from the
renormalized couplings
$l_{i,r}^{\mput{MS}}(\mu)$ and $z_r^{\mput{MS}}(\mu)$
appropriate finite pieces \cite{BBDM}. The procedure is
based on the relation
\bea
\frac{\Gamma(1-w)}{(4\pi)^w} \exp w[\Gamma'(1)+\ln 4\pi]=
\exp {\sum_{n=2}^{\infty}\frac{\zeta ( n ) }
{n}w^n}\co\label{eqmsbar}
\eea
which shows that a simple factor is responsible for these terms.
To remove them, one
 pulls out  a factor
$c^{2w}$ in the definition of the renormalized
couplings in (\ref{eqexpl}),
\bea
l_i(d)&=&\frac{(\mu c)^{2w}}{(4\pi)^2}\left\{\frac{\gamma_i}{2w}
+l_{i,r}^c(\mu,w)\right\}\co\label{eqlmsbar}
\eea
as a result of which $Q_{\mput{loop}}$ in (\ref{eqqloop}) becomes
\bea
Q_{\mput{loop}}=(\mu c)^{4w}\left\{(\mu c)^{-4w}\dot{T}_M^2\,x(d)-
(\mu c)^{-2w}\frac{\dot{T}_M}{(4\pi)^2}\sum_{i=1}^
{4}\left[\frac{\gamma_i}{2w}+l_{i,r}^c(\mu,w)\right]y_i(d)
\right\}
{\hskip-2mm}\per\label{eqqc}
\eea
Pulling out in the analogous manner $c^{4w}$ in Eq. (\ref{eqzd}),
\bea
z(d)=\frac{(\mu c)^{4w}}{(4\pi)^4}\left\{ \cdots
+(4\pi)^4z_r^c(\mu)\right\}\co\label{eqzdmsbar}
\eea
shows that these redefinitions  amount to the change
\bea
\ln \frac{M}{\mu}\rightarrow \ln \frac{M}{\mu}-\ln c\co\nn
\left(l_{i,r}^{\mput{MS}}(\mu),z_r^{\mput{MS}}(\mu)\right)
\rightarrow
\left(l_{i,r}^c(\mu,0),z_r^c(\mu)\right)\label{eqmsmsbar}\co \eea
in  Eq. (\ref{eqqms}).

The traditional choice for $c$ in CHPT is \cite{glann}
 \bea
\ln c=-\frac{1}{2}\left[\ln 4\pi +\Gamma'(1)+1\right]\per\pnn
\eea
In this scheme, one uses the notation \cite{glann}
\bea
l_i(d)=l_i^r(\mu)
+ \frac{\mu^{2w}\gamma_i}{(4\pi)^2}\left\{\frac{1}{2w
} - \frac{ 1 }{ 2}[\ln 4\pi +\Gamma'(1)+1]\right\}
+O(w)\co\pnn
\eea
where
\bea
l_i^r(\mu)=\frac{1}{(4\pi)^2}l_{i,r}^c(\mu,0)\pnn
\eea
according to (\ref{eqlmsbar}). Similarly, we write
\bea
z^r(\mu)\doteq z_r^c(\mu)
\eea
for the above choice of the constant $c$.

We have thus arrived at the final expressions that will be used
for $M_\pi,F_\pi$ and $A(s,t,u)$:
\bea
Q&=&\frac{1}{(4\pi)^4}\left\{x_0\left[1+2 \ln
\frac{M}{\mu}\right]^2+\left[2x_1-\frac{1}{2}\gamma_iy_{i1}-(4\pi
)^2l_i^r(\mu)y_{i0}\right]\left(1+2\ln\frac
{M}{\mu}\right)\right.\nn
&&\left.+x_2-\frac{1}{2}\gamma_iy_{i2}-(4\pi)^2l_i^r(\mu)y_{i1}+(
4\pi)^4 z^r(\mu)\right\}\per\label{eqqfinal}
\eea
The scale dependence of the renormalized couplings is
\bea
\mu\frac{dz^r(\mu)}{d\mu}&=&\frac{2}{(4\pi)^4}[2x_1-(4\pi)^2l_i^r
(\mu)y_{i0}-\gamma_iy_{i1}]\co\nn
\mu\frac{dl_i^r(\mu)}{d\mu}&=&-\frac{\gamma_i}{(4\pi)^2}\per\pnn
\eea

\subsubsection{EOM and Laurent expansion of the coupling constants $l_i$}
We include here a brief discussion of some technical aspects of the
renormalization program at $O(p^6)$. In Sect. 2, it was shown that the
counterterms of $\phi^4$ theory can be written  in different forms
using the EOM. Transforming from one set of counterterms to another leaves
the functional $Z_1$ unchanged and produces an additional local action of
$O(\hbar^2)$ that can always be absorbed by changing the coefficients
of the counterterms in $Z_2$.

The situation in CHPT is similar. Using the EOM of CHPT, both the lagrangians
of $O(p^4)$ and $O(p^6)$ can take different forms. Since to $O(p^6)$
the lagrangian $\cL_6$ enters only at the classical solution, such a
modification
has no effect at all as far as $\cL_6$ is concerned. The same is true for all
quantities of $O(p^4)$. However, the couplings in the lagrangian $\cL_4$ also
appear in one--loop and tree diagrams contributing at $O(p^6)$. In this case,
different forms of $\cL_4$ lead to different results a priori.

As for the $\phi^4$ theory in Sect. 2,
the natural place to discuss those modifications is the generating functional
of Green functions. Referring to a forthcoming paper \cite{Betal97}
for an explicit proof, we present here only
the final result. Lagrangians of $O(p^4)$ that differ by the
EOM lead to generating functionals that differ by local actions of $O(p^6)$.
Since  those actions are of course chirally symmetric
they can always be absorbed by
a redefinition of the low--energy constants of $O(p^6)$.

If the low--energy constants are determined phenomenologically by comparison
with experiment, the result is of course independent of the form of $\cL_4$.
On the other hand, using a model for the constants of $O(p^6)$ like resonance
saturation does not specify the form of $\cL_4$. Therefore, different
forms of $\cL_4$ lead to different numerical results because the constants
of $O(p^6)$ are by definition unchanged. For the $\pi\pi$ scattering amplitude,
it turns out that the coefficients $b_1$,\dots,$b_4$ are affected by this
ambiguity while $b_5$ and $b_6$ are not modified. We come back in Sect. 5
to a numerical discussion of this ambiguity.

The second issue we want to address is the dependence of the
low--energy constants $l_i$ of $O(p^4)$ on the dimension of space--time.
In Eq.~(\ref{eqexpl}), we have defined
functions $l_{i,r}^{\mput{MS}}(\mu,w)$ that contain all non--singular
pieces of the $l_i(d)$. These functions enter in the quantity $Q_1$
in $Q_{\mput{loop}}$ and they are subtracted by
the appropriate decomposition of
$Q_{\mput{tree}}$ in (\ref{eqzd}). As a consequence, only the coefficients
$l_{i,r}^{\mput{MS}}(\mu)\doteq l_{i,r}^{\mput{MS}}(\mu,0)$ appear in
renormalized quantities.

Although this is a completely legal subtraction procedure,
it is by no means obvious that this procedure can be applied
consistently for all possible processes. In other words,
the question is whether
the low--energy constants of $O(p^6)$ can be decomposed as in Eq.~(\ref{eqzd})
in a process--independent way such that only the $l_{i,r}^{\mput{MS}}(\mu)$
appear in all observable quantities. To discuss this issue for the case of
minimal subtraction, we expand the $l_i(d)$ up to first order in $w$:
\beq
l_i(d) = \dfrac{\mu^{2w}}{(4\pi)^2} \left[
\frac{\gamma_i}{2w} + l_{i,r}^{\mput{MS}}(\mu) + \delta_i(\mu)w +
O(w^2)\right]. \label{eqLli}
\eeq
The terms of $O(w^2)$ cannot contribute to $O(p^6)$ because the $l_i$ appear
only in diagrams that are not more singular than $1/w$ (except
for the divergent
parts of the $l_i$ themselves). However, the coefficients
$\delta_i(\mu)$ certainly do appear at $O(p^6)$. The subtraction procedure
(\ref{eqzd}) leads to the following two questions:
\begin{itemize}
\item Do the coefficients $\delta_i(\mu)$ appear always in local
contributions of
$O(p^6)$ only?
\item If yes, can they be absorbed into the coupling constants of $O(p^6)$ in
a process--independent manner?
\end{itemize}

The crucial question is of course the first one. It can be shown
\cite{EM96,Betal97} that the terms involving the $\delta_i$ indeed
take the form of local actions in the generating functional of $O(p^6)$,
see also the analogous discussion in the framework of the
$N$--component $\phi^4$ theory
in Sect. 2.
Although these contributions
depend on the choice of the lagrangian of $O(p^4)$ (EOM),
they can therefore always be absorbed in the
low--energy constants of $O(p^6)$ in a process--independent way. Thus, the
subtraction procedure (\ref{eqzd}) is a special
case of a general method. It has already been used in \cite{BGS94,Bur96,BT97}.

\setcounter{equation}{0}
\section{Elastic $\pi\pi$ scattering to two loops}
\subsection{Pion mass}
\begin{figure}[t]
\begin{center}
\mbox{\epsfxsize=4cm \epsfbox{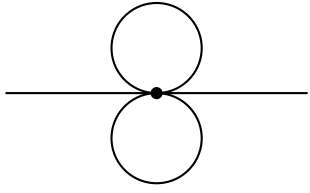}
     }
\caption{Butterfly diagram contribution to the two--point
function (\protect\ref{eqtwopoint}).} \label{f6}
\end{center}
\end{figure}

We consider the connected two--point function
\bea\label{eqtwopoint}
\triangle^{ik}(p^2)=i\int d^dx\,
e^{-ip(x-y)}\la 0|T
{\phi}^i(x){\phi}^k(y)|0\ra_{\mbox\small{conn}} \eea
at two--loop order.
According to the formula (\ref{eqcount}),
the topology of the
graphs is the same as in $O(N)$ theory, except for the
term proportional to
$\mbox{\boldmath$\phi$}^{6}$.
We may therefore again simplify
the calculation
by evaluating the tree and tadpole diagrams displayed in Fig.
 \ref{f3}, with an appropriate choice for the propagators
and for the interaction vertex, and to add
 the sunset diagram Fig. \ref{f2}d$_{10}$,
using the vertices that correspond to ${\cal L}_2$. Finally, the
term from $a_2$ generates  the butterfly diagram Fig.
 \ref{f6}.
 Collecting these contributions, we find
\bea\label{eqzphi}
\triangle^{ik}(p^2)=\left\{\frac{Z}{M_\pi^2-p^2}+
R(p^2)\right\}\delta^{ik}\scs
\eea
where
\bea
M_\pi^2\!\!\!&=&\!\!\!M^2
\!+\!\frac{M^4}{F^2}\left\{2l_3+\frac{T_M}{2M^2}\right\}
+\frac{M^6} { F^4
}\dot{T}_M\left\{Q^M\dot{T}_M-\sum_{i=1}^{3}
Q_i^M\,l_i\right\} +\frac{M^6\,r_M}{F^6}+O(M^8),\nn
Z\!\!\!&=&\!\!\!1-\frac{T_M}{F^2}+
 \frac{M^4}{F^4}\dot{T}_M\left\{Q^Z\dot{T}_M-\sum_{i=1}^{3}
Q_i^Z\, l_i \right\} +\frac{M^4\,r_Z}{F^4}+ O (M^6)\co
\eea
where \cite{Bur96}
\bea
Q^M&=&\frac{1}{96}\left\{204-632w+1263w^2\right\} +O(w^3)\co\nn
Q_1^M&=&\frac{1}{2}\left\{28 -30w+31w^2\right\} +O(w^3)\co\nn
Q_2^M&=&8-10w+11w^2+O(w^3)\co\nn
Q_3^M&=&3-4w+4w^2+O(w^3)\co
\eea
and \cite{Bur96}
\bea
Q^Z&=&\frac{1}{96}\left\{96-464w+1185w^2\right\} +O(w^3)\co\nn
Q_1^Z&=&Q_1^M\scs Q_2^Z=Q_2^M\scs Q_3^Z=2\per
\eea
The function $R(p^2)$ in the propagator is regular at
$p^2=M_\pi^2$. It is not needed in the following, and we no not
display its explicit form \cite{Bur96}
here. The quantities $r_M$ and $r_Z$ denote the
counterterm contributions from the lagrangian ${\cal L}_6$. We note that $r_M$
renders the
pion mass finite at two--loop order.

\subsection{Pion decay constant}

We consider the correlators of two pseudoscalar currents
\cite{glann},
 \bea
G^{ik}(p^2)&=&i\int d^dx\,
e^{-ip(x-y)}\la 0|T
\bar{q}_xi\gamma_5\tau^iq_x\bar{q}_yi\gamma_5\tau^kq_y|0
\ra_{\mbox\small{conn}}\co\nn
\bar{q}&=&(\bar{u},\bar{d})
\eea
 at two--loop order. Apart from the overall normalization,
$G^{ik}$ differs
{}from the two--point function $\triangle^{ik}$ only by terms
generated by the low--energy constants $l_3$ and $l_4$ (and,
of course, by the counterterms from ${\cal L}_6$). We find
\bea
G^{ik}(p^2)=\left\{\frac{G_\pi^2}{M_\pi^2-p^2}+
R_\pi(p^2)\right\}\delta^{ik}\scs
\eea
where
\bea
G_\pi^2&=&G^2\left[1+\frac{M^2}{F^2}\left\{2l_4-4l_3-
\frac{T_M}{M^2}
\right\} +
\frac{M^4}{F^4}\dot{T}_M\left\{Q^G\dot{T}_M-\sum_{i=1}^{4}
Q_i^G l_i \right\}\right.\nn
&&+\left.
\frac{M^4}{F^4}\left\{l_4^2+8l_4l_3+4l_3^2+r_G\right\}\right]
+O(M^6)\co
\eea
with
\bea
G&=&2FB\co\nn
Q^G&=&Q^Z\scs Q_1^G=Q_1^Z\scs Q_2^G=Q_2^Z\co\nn
Q_3^G&=&2(8-7w+7w^2)+O(w^3)\co\nn
Q_4^G&=&1-w+w^2+O(w^3)\per
\eea
The counterterm contribution from ${\cal L}_6$ is denoted by $r_G$. It renders
$G_\pi$ finite at two--loop order. The regular part $R_\pi$ is not needed
in the following, and we therefore do not display it here.

 The relation between the divergence of the axial current and the
pseudoscalar density implies
\bea
F_\pi M_\pi^2=\hat{m}G_\pi\per
\eea
The above expressions for the pion mass and for the residue
$G_\pi$ therefore
allow us to calculate the pion decay constant to two loops:
 \bea
F_\pi=F\left[1\!+\! \frac{M^2}{F^2}\left\{l_4\!-\!\frac{T_M}{F^2}\right\}
+\frac{M^4}{F^4}\left\{r_F \!+\!\dot{T}_M\left(Q^F\dot{T}_M
-\sum_{i=1}^{4}Q_i^Fl_i\right)\right\}\right] + O(M^6)\co\nn
\eea
with
\bea
Q^F&=&-\frac{1}{192}(240 -656w+1125w^2)+O(w^3)\co\nn
Q_1^F&=&-\frac{1}{2}Q_1^G\scs Q_2^F=-\frac{1}{2}Q_2^G\scs
Q_3^F=2\scs Q_4^F=\frac{1}{2}Q_4^G\per
\eea
The counterterm contribution from ${\cal L}_6$ is denoted by $r_F$. It renders
$F_\pi$ finite at two--loop order.

\subsection{The $\pi\pi$ scattering amplitude}

We have discussed in Sect. 2 how one can simplify the
calculation of the elastic scattering amplitude in the
$N$--component $\phi^4$ theory
by using the renormalized one--loop amplitude as a single nonlocal vertex.
In case of CHPT, there are a few modifications to be taken
into account:
\begin{enumerate}
\item
As we discussed already in the case of the pion mass and pion
decay constant,  there are additional diagrams to be taken
into account, see the formula
(\ref{eqcount}).  Some are displayed in
Fig. \ref{f7}.
\begin{figure}[t]
\begin{center}
\mbox{\epsfxsize=12cm \epsfbox{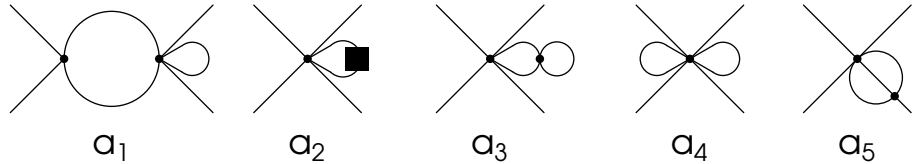}
     }
\caption{Contributions to elastic
$\pi\pi$ scattering in CHPT. The graphs displayed have a different
topology than the ones occurring in $\phi^4$ theory. The filled box denotes a
contribution proportional to the low--energy constant $l_3$. Graph a$_1$
is included by use of the formula indicated in Fig.
\protect\ref{f5}, whereas graphs a$_2\cdots$ a$_5$ are not.
These
must be taken into account separately.\label{f7}} \end{center}
\end{figure}
The first of these is taken care of by
using  the formula illustrated in Fig. \ref{f5}, whereas
the tadpole (Fig. \ref{f7}a$_2$), the
 butterfly (Figs. \ref{f7}a$_3$,a$_4$) and
sunset diagram (Fig. \ref{f7}a$_5$)
have to be added by hand.
\item
Since the effective lagrangian in CHPT has derivative
interactions, the vertices entering the calculation
have a nontrivial off--shell behaviour. This means that the
one--loop amplitude entering the above formulae must first
be calculated with  the appropriate off--shell legs in $d$
dimensions. Since
the relevant expression is not available in the literature, we
display it in appendix \ref{appipi}.
 \end{enumerate}
Needless to say that, despite the labour saving organization of
the calculation
that we just described, it is still rather long and tedious.

In
analogy to what we have presented above, we might now display the
 amplitude  in $d$ dimensions.
 Since
the expression is rather long, we prefer to display only the
renormalized final expression.

 We use the notation
\bea
&&\langle \pi^d(p_4)\pi^c(p_3)\;\mbox{out}|\pi^a(p_1) \pi^b(p_2)\;\mbox{in}
\rangle =
\langle \pi^d(p_4)\pi^c(p_3)\;\mbox{in}|\pi^a(p_1) \pi^b(p_2)\;\mbox{in}\ra \nn
&&\hspace{3cm}+i
(2\pi)^4\delta^{4}(P_f-P_i)\left\{\delta^{ab}\delta^{cd}A(s,t,u)
+\mbox{cycl.} \right\}\co\nonumber
\eea
where $s,t,u$ are the usual Mandelstam variables, expressed in units of the
physical pion mass squared $M_\pi^2$,
\bea
s&=&(p_1+p_2)^2/M_\pi^2 \scs t=(p_3-p_1)^2/M_\pi^2 \scs
u=(p_4-p_1)^2/M_\pi^2\per\label{eqmandel}
\eea
Using these dimensionless quantities, the momentum expansion of the amplitude
amounts to a Taylor series in
\bea
{\it x_2}=\frac{M_\pi^2}{F_\pi^2}\co\nonumber
\eea
where $F_\pi$ denotes the physical pion decay constant.
We find
 \bea A(s,t,u)&=& \hspace{.3cm}{\it
x_2}\left[s-1\right]\nn
&&+{\it x_2}^2\left[b_1+b _ 2s + b_3 s^2 +
b_4 ( t - u )^2\right]\nn
&&+{\it x_2}^2\left[F^{(1)}(s) +G^{(1)}(s,t)+G^{(1)}(s,u)\right]\nn
&&+{\it x_2}^3\left[b_5s^3+b_6s(t-u)^2\right]\nn
&&+{\it x_2}^3\left[F^{(2)}(s)+G^{(2)}(s,t)
+G^{(2)}(s,u)\right]\,\!\nn
&&+O({\it x_2}^4)\co
\label{amptot}
 \eea
with
 \begin{eqnarray}
{ F^{(1)}}(\,{s}\,) &=& {\displaystyle \frac {1}{2}}\,
 \bar{J}(\,{s}\,)\,(\,{s}^{2} - 1\,)\co\nn
{ G^{(1)}}(\,{s}, {t}) &=& {\displaystyle \frac {1}{6}}\,
 \bar{J}(\,{t}\,)\,(14\,-\,4\, s\,-\,10\, t\,+\,s\, t\,+\,2\, t^2\,)\co\nn
{ F^{(2)}}(\,{s}\,) &=& \bar{J}(\,{s}\,) \left\{ {\vrule
height0.79em width0em depth0.79em} \right. \! \frac{1}{16 \pi^2}
\left(\! {\displaystyle
\frac {503}{108}}\,{s}^{3} - {\displaystyle \frac {929}{54}}\,{s}
^{2} + {\displaystyle \frac {887}{27}}\,{s} - {\displaystyle
\frac {140}{9}} \right) \nn
 &+&  {b_1}\,(\, 4\,{s}\, - 3) + {b_2}\,(\,
 {s}^{2} + 4\,{s}\, - 4) \nn
 &+& {\displaystyle \frac {{b_3}}{3}}\,\,(\,8\,{s}
^{3} - 21\,{s}^{2} + 48\,{s} - 32\,)
 + {\displaystyle \frac {{b_4}}{3}}\,
(\,16\,{s}^{3} - 71\,{s}^{2} + 112\,{s} - 48\,)
 \! \left. {\vrule height0.79em width0em depth0.79em}
 \right\} \nn
 &+ &\mbox{} {\displaystyle \frac {1}{18}}\,{ K_1}(\,{s}
\,)\, \left\{ \! \! \,20\,{s}^{3} - 119\,{s}^{2} + 210\,{s} - 135 -
{\displaystyle \frac {9}{16}}\,{ \pi}^{2}\,(\,{s} - 4\,)\, \!  \right\}  \nn
 &+ & \mbox{} {\displaystyle \frac {1}{32}}\,{ K_2}(\,{s}\,)\,
 \left\{ \! \,{s}\,{ \pi}^{2} - 24\, \!  \right\}  +
{\displaystyle \frac {1}{9}}\,{ K_3}(\,{s}\,)\,\left\{\,3\,{s}^{2} -
17\,{s} + 9\,\right\}\co \nn
{ G^{(2)}}(\,{s}, {t}\,) &=& \bar{J}(\,{t}\,) \left\{
{\vrule height0.79em width0em depth0.79em} \right. \!
\frac{1}{16 \pi^2}
\left[
{\displaystyle \frac {412}{27}}\! -\! {\displaystyle \frac {s}{54}}
({t}^{2} + 5\,{t} + 159)
\! -\! t \left(\frac{267}{216}{t}^{2} - \frac{727}{108}{t} +
\frac{1571}{108} \right) \right] \nn
 &+&   {b_1}\,(2
 - {t})
+ {\displaystyle \frac {{b_2}}{3}}({t} - 4
)(2\,{t} + {s} - 5)
- {\displaystyle \frac {{b_3}}{6}}
({t} - 4)^{2}(3{t} + 2{s} - 8) \nn
&+& {\displaystyle \frac {{b_4}}{6}}\left(2{s}
(3{t} - 4)({t} - 4) - 32 t + 40t^2 - 11{t}^{3}\,\right)\!
  \left. {\vrule
height0.79em width0em depth0.79em} \right\} \mbox{} \nn
 & +&
{\displaystyle \frac {1}{36}}{ K_1}(\,{t}\,)
\left\{\,174 + 8\,{s} - 10\,{t}^{3}
 + 72\,{t}^{2} - 185\,{t} - {\displaystyle \frac {{\pi}^2}{16}}\,
(\,{t} - 4\,)\,(\,3\,{s}\! -\! 8\,)\, \!
 \right\}  \nn
 &+ & \mbox{} {\displaystyle \frac {1}{9}}\,{ K_2}(\,{t}\,)
\, \left\{ \! \,1 + 4\,{s} + {\displaystyle \frac {{\pi}^2}{64}}
\,{t}\,(\,3\,{s} - 8\,)\
\, \!  \right\}  \nn
 &+& {\displaystyle \frac {1}{9}}\,{ K_3}(\,{t}\,)\left\{
1 + 3{s}{t} - {s} + 3{t}^{2} - 9{t}\right\}
+ {\displaystyle \frac {5}{3}}\,{ K_4}(\,{t}\,)\,\left\{\,4 - 2\,{s} -
{t}\,\right\}\per
\label{amptot1}
\end{eqnarray}
The loop functions $\bar{J}$ and $K_i$ are
\begin{eqnarray*}
\left(\begin{array}{l}\bar{J}\\
K_1\\
K_2\\
K_3\\
\end{array}\right)
=
\left(\begin{array}{cccc}
0&0&z&-4N\\
0&z&0&0\\
0&z^2&0&8\\
Nzs^{-1}&0&\pi^2(Ns)^{-1}&\pi^2\\
 \end{array} \right)
 \left(\begin{array}{c}
{h}^3\\
{h}^2\\
{h}\\
\displaystyle{-(2N^2)^{-1}}
\end{array}
\right)\co
\end{eqnarray*}
and
\begin{eqnarray*}
K_4&=&\frac{1}{sz}\left(\frac{1}{2}K_1+\frac{1}{3}K_3+
\frac{1}{N}\bar{J}
+\frac{(\pi^2-6)s}{12N^2}\right)\co \nn
\end{eqnarray*}
where
\[
{h}(s)=\frac{1}{N\sqrt{z}}\ln
\frac{\sqrt{z}-1}{\sqrt{z}+1} \quad ,\qquad z=1-\frac{4}{s} \; , \;
N=16\pi^2\per
\]
The functions $s^{-1}\bar{J}$ and $s^{-1}K_i$ are analytic in the complex
$s$--plane (cut along
the positive real axis for $s \geq 4$), and they vanish
 as $|s|$ tends to infinity. Their real and
imaginary parts are continuous
at $s=4$.
The coefficients $b_i$ in the polynomial part
are given in  appendix \ref{apbi}.
\vspace{.5cm}

Finally, we compare our result with the calculation performed
 in Ref. \cite{KMSF95}. There are
two basic differences
between the two works. First, the present
calculation is
done in conventional chiral perturbation theory \cite{glann},
whereas the authors
of \cite{KMSF95} work in a scenario in which the quark condensate
may
be small or zero \cite{gchpt} (generalized chiral perturbation
theory). The
second
difference concerns the use of a lagrangian framework in the
present approach, whereas S--matrix methods are applied in
\cite{KMSF95}: Starting from the expression for the amplitude at
$O(p^4)$, unitarity allows one to determine the
absorptive part at $O(p^6)$.  One can then construct an
amplitude  that has exactly these absorptive
parts.
Of course, this procedure does not determine the polynomial
part at this order. To illustrate, the diagrams of  Fig.
 \ref{f7}a$_{2,3,4,5}$ do not have an absorptive part and can
therefore
not be determined in this manner. There are additional graphs
of this kind. In the amplitude given in  \cite{KMSF95}, these
diagrams -- and further polynomial contributions from graphs that
do have an absorptive part --
are lumped into a polynomial
\bea
a_1+a_2s+a_3s^2+a_4(t-u)^2+a_5s^3+a_6s(t-u)^2\label{eqgch}
\eea
with a priori unknown coefficients that depend on the pion
mass and on the low--energy constants.

With this procedure, one does not have to work out individual
Feynman diagrams, because only the total absorptive part at $O(p^6)$
 -- fixed through unitarity by the amplitude at $O(p^4)$ -- is needed.
In addition, the expansion of the pion mass, of the
pion decay constant and of the wave function renormalization
need not be worked out at  $O(p^6)$.

On the other hand, our result (\ref{amptot}, \ref{amptot1}) at
$O(p^6)$ -- obtained by painstakingly evaluating in the manner
described above all the
Feynman diagrams generated by the effective lagrangian (\ref{eqstring}) --
reveals  in addition to the findings of \cite{KMSF95} the dependence
of the six coefficients $a_i$ (\ref{eqgch}) on the pion mass
 and on the low--energy constants of both
$O(p^4)$ and $O(p^6)$, see appendix \ref{apbi}. In particular,
 we find that these coefficients contain mass singularities that
 are known to be important numerically
both at $O(p^4)$ \cite{glpl,glann} and at $O(p^6)$
\cite{colangelo,BCEGS1} (see also
Refs. \cite{BGS94,Bur96,BT97}). Furthermore, the knowledge of the
mass dependence of the amplitude allows one to evaluate
all quantities   even at unphysical
values of the quark mass, and to confront the
result with  lattice calculations \cite{colangelolat}. This
is not possible with the amplitude provided in Ref. \cite{KMSF95}.

We have checked that
the amplitude of \cite{KMSF95}
(restricted to standard CHPT) and
the field theoretic calculation presented here agree as far as
the absorptive part of the amplitude and the general structure of
the real part is concerned.

\setcounter{equation}{0}

\section{Numerical analysis}

 The elastic $\pi\pi$ scattering amplitude is expressed in terms of the
external momenta, the physical pion mass, the physical pion decay constant,
and the coefficients $b_i$,
\bea
A(s,t,u)=f(p_1,\ldots,p_4; M_\pi,F_\pi; b_1,\ldots,b_6)\per\pnn
\eea
 Before one can perform a numerical analysis,
 one thus needs an estimate of the low--energy couplings  $b_i$.
The key relations (\ref{eqb_i})  used in \cite{BCEGS1} for that purpose
fix the $b_i$ in terms of
 \begin{itemize} \item[--]  chiral
logarithms $\displaystyle{L=\frac{1}{16\pi^2}\ln{\frac{M_\pi^2}{\mu^2}}}\co$
\item[--]
 the low--energy couplings
$l_1^r(\mu),\ldots,l_4^r(\mu)$ from ${\cal L}_4\co$
\item[--]
 the low--energy couplings
$r_1^r(\mu),\ldots,r_6^r(\mu)$ generated by
${\cal L}_6\per$
\end{itemize}
For a given set of the low--energy constants $l_i^r$ it,
therefore, suffices to estimate the new couplings $r_i^r$ at
$O(p^6)$. To achieve an order--of--magnitude estimate, we use a method
that has been successfully tested at $O(p^4)$ \cite{glann,EGPR89}:
 We work out the contributions
 from  exchanges of
heavy states to the scattering amplitude and assume that these
effects account
for the bulk part in the low--energy couplings at
$O(p^6)$.
In particular, we include the effect of meson resonance exchange with
masses smaller than 1 GeV,
i.e., vector and scalar resonances. In addition, we
also consider kaon and eta contributions of $O(p^6)$. We
describe in the following subsection the
 relevant resonance couplings and work out the corresponding
values for the $r_i$.

\subsection{Resonance saturation}
\subsubsection{Vector meson exchange}
At $O(p^4)$, the contributions of vector meson resonances to the low--energy
constants are obtained most naturally with the tensor field representation
of spin-1 mesons \cite{glann,EGLPR89}. The situation is different at $O(p^6)$
where the more common vector field formalism produces directly the relevant
counterterm couplings, as observed previously for
Compton scattering on pions and for $\gamma\gamma \to \pi \pi$ \cite{EPR90,
BGS94}. We follow the same procedure here and comment later on the difference
to the tensor field representation.

The relevant couplings of the vector field $\hat V_\mu$, representing the octet
of vector mesons (the singlet does not contribute to
$\pi\pi$ scattering), to the
pseudoscalar mesons are given by \cite{EGLPR89}
\beq
\cL_V = - \frac{i g_V}{2 \sqrt{2}} \bla \hat V_{\mu\nu} [u^\mu, u^\nu]
\bra + f_\chi \bla \hat V_\mu [u^\mu, \chi_-] \bra \label{eqVM}
\eeq
$$
\hat V_{\mu\nu} = \nabla_\mu \hat V_\nu - \nabla_\nu \hat V_\mu
$$
with real dimensionless coupling constants $g_V, f_\chi$. We switch
to chiral $SU(3)$ here (and also in the next subsection for the discussion
of scalar exchange) to determine both couplings $g_V, f_\chi$ from vector
meson decays. From the Lagrangian (\ref{eqVM}) one finds that the
combination
$$
g_V + 2 \sqrt{2} f_\chi \dfrac{M_1^2 + M_2^2}{M_V^2}
$$
determines the amplitude for the decay of a vector meson  into two
pseudoscalars. From the experimental widths for $\rho \to \pi\pi$ and
$K^* \to K \pi$ one obtains $g_V f_\chi< 0$ with
\beq
g_V = 0.09~, \qquad f_\chi = - 0.03~,
\eeq
if we define $g_V$ to be positive. $V$ exchange on the
basis of (\ref{eqVM}) gives rise to three types of contributions proportional
to $g_V^2$, $g_V f_\chi$ and $f_\chi^2$, respectively. With
\beqan
a_V &=& \dfrac{g_V^2 F_\pi^2}{M_V^2}= 1.2\cdot 10^{-4} \\
b_V &=& \dfrac{4 \sqrt{2} g_V  f_\chi F_\pi^2}{M_V^2}=  - 2.2\cdot 10^{-4} \\
c_V &=& \dfrac{32 f_\chi^2 F_\pi^2}{M_V^2}= 4.2\cdot 10^{-4}~,
\eeqan
using $M_V=M_\rho$ for the numerical values, we obtain the following $V$
exchange contributions to the low--energy constants $r_i$:
\beqa
r_1^V &=& -16 a_V -16 b_V -4 c_V \nn
r_2^V &=& 20 a_V +16 b_V +3 c_V \nn
r_3^V &=& -7 a_V -3 b_V  \label{eqriV} \\
r_4^V &=& a_V + b_V \nn
r_5^V &=& \frac{3}{4} a_V  \nn
r_6^V &=& \frac{1}{4} a_V \no ~.
\eeqa

In the tensor field representation, there is a single coupling (relevant for
$\pi\pi$ scattering) between vector
mesons and pseudoscalars at lowest order \cite{glann,EGPR89}.
Up to normalization,
the corresponding coupling constant is given by $g_V$ . For instance,
expanding the $\rho$ exchange amplitude of Ref.~\cite{glann} to
$O(p^6)$ is equivalent to (\ref{eqriV}) with $b_V = c_V = 0$.
Higher--order couplings can only contribute
at $O(p^6)$ through interference with the $g_V$ amplitude. Thus, the
contributions involving $f_\chi^2$ that appear naturally in the
vector field formalism would have to be added by hand in the tensor field
representation as explicit local counterterms.

\subsubsection{Scalar meson exchange}
To estimate the effect of scalar resonances for the $r_i$, we take the
lowest--order lagrangian of Ref.~\cite{EGPR89} for octet and singlet scalar
fields $S$, $S_1$,
\beq
\cL_S = c_d \bla S u^\mu u_\mu \bra + c_m \bla S \chi_+ \bra
+ \wt{c_d} S_1 \bla u^\mu u_\mu \bra + \wt{c_m} S_1 \bla \chi_+ \bra~,
\eeq
and expand the resonance exchange amplitudes up to $O(p^6)$. The resulting
scalar contributions to the $r_i$ are:
\beqa
r_1^S &=& 0 \nn
r_2^S &=& \frac{8 F_\pi^2}{3 M_S^4}(c_m - c_d)^2 + \frac{16 F_\pi^2}
{M_{S_1}^4}(\wt{c_m} - \wt{c_d})^2 \nn
r_3^S &=& \frac{8 F_\pi^2}{3 M_S^4}c_d(c_m - c_d) + \frac{16 F_\pi^2}
{M_{S_1}^4}\wt{c_d}(\wt{c_m} - \wt{c_d}) \nn
r_4^S &=& 0 \\
r_5^S &=& \frac{2 F_\pi^2}{3 M_S^4}c_d^2 + \frac{4 F_\pi^2}
{M_{S_1}^4}\wt{c_d}^2 \nn
r_6^S &=& 0~.\no
\eeqa
Note that the physical pion mass and the physical pion
decay constant receive contributions from scalar exchange as
well.

For the numerical evaluation we use the same values for masses and coupling
constants as in Ref.~\cite{EGPR89}:
\beqan
M_S = M_{S_1} &=& 983 \mbox{\rm ~MeV} \\
c_m = 42 \mbox{\rm ~MeV}~,& &  c_d = 32 \mbox{\rm ~MeV}\\
\wt{c_m} = c_m/\sqrt{3}~,& &  \wt{c_d} = c_d/\sqrt{3}~.
\eeqan

\subsubsection{Kaon and eta contributions}
In the framework of chiral $SU(3)$, $K$ and $\eta$ mesons contribute at
$O(p^4)$ to $\pi\pi$ scattering via loop diagrams. Restricting the scattering
amplitude of $O(p^4)$ evaluated in chiral $SU(3)$ \cite{KMSF95} to
$SU(2)\times SU(2)$ by an expansion in inverse powers of the strange quark
mass, one arrives at the following contributions to the $r_i$ due to $K$ and
$\eta$ mesons:
\beqa
r_1^K &=& \dfrac{31 F_\pi^2}{5760 \pi^2 M_K^2} \nn
r_2^K &=& - \dfrac{11 F_\pi^2}{2304 \pi^2 M_K^2} \nn
r_3^K &=& - \dfrac{29 F_\pi^2}{7680 \pi^2 M_K^2} \nn
r_4^K &=& - \dfrac{F_\pi^2}{2560 \pi^2 M_K^2} \\
r_5^K &=& \dfrac{23 F_\pi^2}{15360 \pi^2 M_K^2} \nn
r_6^K &=& \dfrac{F_\pi^2}{15360 \pi^2 M_K^2} ~.\no
\eeqa

\subsubsection{Numerical values of the $r_i$}
Putting the various contributions of the previous subsections together, we
obtain the following numerical estimates for what we shall call the resonance
contributions to the low--energy constants of $O(p^6)$:
\bea\label{eqriR}
r_1^R &=& - 0.6 \cdot 10^{-4}\nn
r_2^R &=& 1.3 \cdot 10^{-4}\nn
r_3^R &=& - 1.7 \cdot 10^{-4}\nn
r_4^R &=& - 1.0 \cdot 10^{-4} \label{eqrir}\\
r_5^R &=& 1.1 \cdot 10^{-4}\nn
r_6^R &=& 0.3 \cdot 10^{-4}\no~.
\eeqa
These values are the ones used in Ref.
\cite{BCEGS1}.
They indicate an $O(p^6)$ version of vector meson
dominance: Despite
some strong cancellations (especially for $r_1^V$, $r_2^V$), all six
constants $r_i^R$ are dominated by the exchange of vector
mesons.\footnote{
 Other estimates of
the low--energy constants of $O(p^6)$ can be found in
Refs.~\cite{rirlit}.}

Of course, the constants $b_i$ become
scale dependent if we now replace the $r_i^r(\mu)$ by the above
$r_i^R$.
At $O(p^4)$, the empirically observed resonance saturation of low--energy
constants amounts to the statement \cite{EGPR89} that the resonance
contributions describe the phenomenologically determined coupling constants
quite well for a scale $\mu$ between 500 MeV and 1 GeV.
Pending a more refined analysis along the lines discussed
below, we assume for the time being that the same approximation is meaningful
also at $O(p^6)$. Of course, only observables that are relatively insensitive
to scale changes in the range mentioned can be  estimated in a
reasonable manner in this way.

\subsection{The constants $b_i$}
 Together with the couplings $l_i^r$ at $O(p^4)$, the
estimates (\ref{eqrir}) allow one to work out the
constants $b_1,\ldots,b_6$.
We used in Ref.~\cite{BCEGS1} values for the
$l_i^r$ that were determined mainly from an $O(p^4)$ analysis
\cite{glann}, with additional input from a dispersive estimate of
higher--order effects in $K_{l4}$ decays \cite{bcgke4}. As emphasized in our
previous paper, all those determinations were faced with the problem that the
$l_i$ are mass independent, whereas the physical quantities, from which the
$l_i^r$ were determined, include quark mass effects. We will
illustrate below the relevance of those quark mass effects by example.

With the scale--independent couplings $\olc{l}{_i}$ (cf. App. D) found
in \cite{glann,bcgke4},
\beqa
\begin{array}{rrrr}
\olc{l}{_1}=&-1.7 \co&
\olc{l}{_2}=& 6.1 \co \\
\olc{l}{_3}=& 2.9\co&
\olc{l}{_4}=& 4.3 \;\; ,
\end{array} \label{eqliold}
\eeqa
and with the constants $r_i^R$ in (\ref{eqrir}),
one arrives\footnote{
As in Ref.~\cite{BCEGS1}, we use $F_\pi=93.2$ MeV,
$M_\pi=139.57$ MeV.}
at  the $b_i$ displayed in table \ref{tab:seti}.
 There, we have split the contributions to the $b_i$ into one--loop and
two--loop effects. Furthermore, we indicate separately the
contributions from the $r_i^R$. The values shown correspond to
 the scale $\mu=1$ GeV ($\mu=500$ MeV in brackets). (We display
the values to two digits for ease of reproduction of the
numbers.) Below we refer to the $b_i$ in this table as set I.
The following remarks are in order.

\begin{table}[ht]
\caption{The constants $b_i$ (set I). We use
Eqs.~(\protect\ref{eqrir},\protect\ref{eqliold}) as input to evaluate
$b_i$ from Eq.~(\protect\ref{eqb_i}). The numbers correspond to
$\mu=1$ GeV, in brackets we display the corresponding values at
$\mu=500$ MeV.} \label{tab:seti}
\begin{center}
\vspace{.5cm}
\begin{tabular}{c||r|r|r|c}  \hline

       & 1-loop&2-loops,  $r_i$=0& from $r_i^R$ & total\\ \hline

  $10^2 b_1$&-7.34&-1.76 (-1.82) & -0.01 & -9.11 (-9.17) \\

  $10^2 b_2$& 6.74& 2.07 ( 2.12) &  0.03 &  8.84 ( 8.89) \\

  $10^3 b_3$&-0.84&-3.05 (-3.43) & -0.38 & -4.27 (-4.65) \\

  $10^3 b_4$& 5.56& 1.72 ( 1.71) & -0.22 &  7.05 ( 7.04) \\

  $10^4 b_5$&     & 1.22 ( 1.24) &  1.10 &  2.32 ( 2.34) \\

  $10^4 b_6$&     & 1.19 ( 0.77) &  0.30 &  1.49 ( 1.07) \\
\hline
\end{tabular}
\end{center}
\end{table}

\begin{enumerate}
\item[i)]
The change induced by $\mu=1$ GeV $\rightarrow \mu=500$ MeV
is of the same order or bigger than the
contributions from $r_i^R$ (see also table \ref{tab:setii}), with the
possible exception of $b_4$.
\item[ii)]
Besides the scale dependence, there is yet another source of
uncertainty that has to do with using the EOM in the
lagrangian of $O(p^4)$. As discussed in Sects.~2 (for $\phi^4$) and 3
(for CHPT), this ambiguity can always be resolved by a redefinition of the
coupling constants of $O(p^6)$ --  physical observables are
 of course unaffected.
Although this is a completely general result for the generating functional to
$O(p^6)$ \cite{Betal97}, we nevertheless face a practical problem here because
our estimates of the $r_i^r$ via resonance exchange are not only scale
independent but they also make no reference to the form of the lagrangian of
$O(p^4)$. We have checked that using
the $O(p^4)$ lagrangians of either Ref.~\cite{glann}  or  of
Ref.~\cite{GSS88} induces a change in  $b_{1,2,3,4}$ by an amount that is
smaller than the change due to a different choice of scale.
 The constants  $b_{5,6}$ are unaffected by this procedure.
 \item[iii)]
The main uncertainties in the $b_i$ stem from the couplings of
$O(p^4)$, as we  now illustrate. In table \ref{tab:bi1loop},
we have displayed the contributions from $\olc{l}{_i}$ to $b_i$ at
one--loop order.
 They are at least an order of magnitude bigger than the
ones from the $r_i^R$ -- therefore, uncertainties in the latter
are swamped by uncertainties in the $\olc{l}{_i}$. One of these are quark
mass effects, mentioned above. To illustrate these,
 we repeat
the analysis
of Ref.~\cite{glann} to fix $\olc{l}{_1}$, $\olc{l}{_2}$ from the $D$--wave
scattering lengths $a_2^0$, $a_2^2$, but now to $O(p^6)$. Keeping
$\olc{l}{_3}$,
$\olc{l}{_4}$ fixed, the experimental values \cite{threshexp} given in table
\ref{tab:thresh} lead to
\beqa
\ba{rr}
\olc{l}{_1} &= -1.5 \\
\olc{l}{_2} &=  4.5
\ea\label{eqlinew}
\eeqa
if $\olc{l}{_3},\olc{l}{_4}$ from (\ref{eqliold}) are used as input.
The $b_i$ that result from this exercise are displayed in table
\ref{tab:setii} and referred to as  set II in the following.
Since $b_{1,2}$ do not depend on $\olc{l}{_2}$, they
remain largely unaffected, whereas the change in
$b_{3,\ldots,6}$ is seen to be substantial.
The values (\ref{eqlinew}) should be compared with the values
obtained at $O(p^4)$, $\olc{l}{_1}=-2.3$ and $\olc{l}{_2}=6.0$,
from the same input~\cite{glann}. Values similar to (\ref{eqlinew}) were
found in \cite{gkms97} comparing a Roy equation fit
to our $p^6$ amplitude~\cite{BCEGS1}.

\begin{table}[ht]
\caption{Contributions from the individual
$\olc{l}{_i}$ in Eq.~(\protect\ref{eqliold}) to the constants $b_i$
in the one--loop approximation. Note that $b_{5,6}$ start at two--loop
order.}\label{tab:bi1loop}
\begin{center}
\vspace{.5cm}
\begin{tabular}{c||r|r|r|r|r|r}  \hline

  1-loop&$\olc{l}{_1}$&$\olc{l}{_2}$&$\olc{l}{_3}$&$\olc{l}{_4}$&analytic&
   total\\ \hline

  $10^2 b_1$&-1.44& 0  & -0.92& -5.44& 0.46 &-7.34 \\

  $10^2 b_2$& 1.44& 0  &  0   &  5.44& -0.14&6.74  \\

  $10^3 b_3$&-3.59&6.44&  0   &  0   &-3.69&-0.84  \\

  $10^3 b_4$&  0  &6.44&  0   &  0   & -0.88&5.56 \\
\hline
\end{tabular}
\end{center}
\end{table}

\begin{table}[ht]
\caption{The constants $b_i$ (set II). We use
Eqs.~(\protect\ref{eqrir},\protect\ref{eqlinew}) as input to evaluate
$b_i$ from Eq.~(\protect\ref{eqb_i}). The numbers correspond to
$\mu=1$ GeV, in brackets we display the corresponding values at
$\mu=500$ MeV.} \label{tab:setii}
\begin{center}
\vspace{.5cm}
\begin{tabular}{c||r|r|r|c}  \hline

       & 1-loop&2-loops, $r_i$=0& from $r_i^R$ & total\\ \hline

  $10^2 b_1$&-7.17&-1.44 (-1.62) & -0.01 & -8.63 (-8.80) \\

  $10^2 b_2$& 6.57& 1.41 ( 1.71) &  0.03 &  8.01 ( 8.31) \\

  $10^3 b_3$&-2.11&-0.14 (-1.85) & -0.38 & -2.63 (-4.34) \\

  $10^3 b_4$& 3.87& 1.17 ( 1.20) & -0.22 &  4.82 ( 4.85) \\

  $10^4 b_5$&     &-1.47 (-0.22) &  1.10 & -0.37 ( 0.88) \\

  $10^4 b_6$&     & 0.53 ( 0.42) &  0.30 &  0.83 ( 0.72) \\
\hline
\end{tabular}
\end{center}
\end{table}
\item[iv)]
Similar effects are expected from changes in
$\olc{l}{_4}$ that does contribute substantially to $b_{1,2}$.
As we already mentioned in our previous work
\cite{BCEGS1}, a more
detailed analysis is therefore needed to obtain reliable values
for the $\olc{l}{_i}$, consistent with an analysis at $O(p^6)$, and, at
the same time,
 values for the  $b_i$ with error bars attached. Such an
analysis is under way \cite{royanalysis}.
 \item[v)]
This brings us to the work of Ref.~\cite{KMSF96}
 who find from
dispersion sum rules the values
\beqa
\left.\ba{rrr}
b_3 =& (-3.7 \pm 2.4)  \cdot 10^{-3} \\
b_4 =& (4.8 \pm 0.3)  \cdot 10^{-3} \\
b_5 =& (1.4 \pm 0.6)  \cdot 10^{-4} \\
b_6 =& (1.0 \pm 0.1)  \cdot 10^{-4}
\ea \right\}\ba{r} \\ \\
 \ea\label{eqbistern}
 \eeqa
Within the error bars quoted, these values are consistent with
our set II.
 The theoretical framework used in \cite{KMSF96} does not allow
these
authors to determine $b_{1,2}$  -- rather, these
quantities can in principle
be determined from precise data on elastic $\pi\pi$ scattering,
allowing then a determination of the size of the quark condensate
in the chiral limit.
 \item[vi)]
Recently, Wanders has also extracted the $b_{3,4,5,6}$ using
crossing--symmetric sum rules \cite{wanders97}.  Taking the $O(p^4)$
expressions for $b_1$, $b_2$ as input, he finds that at $O(p^6)$
three of the $b_i$ are only weakly sensitive to the choice of the energy
separating high-- and low--energy components in the sum rules. The resulting
values (for the interpretation of the errors we refer to \cite{wanders97})
\beqa
\left.\ba{rrr}
b_3 =& (-2.55 \pm 0.20)  \cdot 10^{-3} \\
b_4 =& (4.55 \pm 0.15)  \cdot 10^{-3} \\
b_6 =& (0.92 \pm 0.03)  \cdot 10^{-4}
\ea \right\}\ba{r} \\ \\
 \ea\label{eqbiwanders}
 \eeqa
are consistent with our set II and with the values (\ref{eqbistern}) of
Knecht et al.~\cite{KMSF96}.
\end{enumerate}

\subsection{Threshold parameters and phase shifts}

To compare the theoretical amplitudes with  data on $\pi\pi$
scattering, one expands the combinations with definite
isospin in the $s$-channel
\bea
T^0(s,t)&=&3A(s,t,u) +A(t,u,s)+A(u,s,t)\nn
T^1(s,t)&=&A(t,u,s)-A(u,s,t)\\ \label{eqiso}
T^2(s,t)&=&A(t,u,s) +A(u,s,t)\nonumber
\eea
into partial waves,
\bea
T^I(s,t)&=&32\pi
\sum_{l=0}^{\infty}(2l+1)P_l(\cos{\theta})t_l^I (s)\per\label{eqpartial}
\eea
Unitarity implies that in the elastic region\footnote{We recall our
normalization of the Mandelstam variables in (\ref{eqmandel}).}
 $4 \le s \le 16$
the partial
wave amplitudes $t_l^I$ are described by real phase shifts $\delta_l^I$,
\beq
t_l^I(s)=\left(\frac{s}{s-4}\right)^{1/2}\frac{1}{2i}
\{e^{2i\delta_l^I(s)}-1\}\per \label{equnitary}
\eeq
The behaviour of the partial waves  near threshold is of the form
\beq
\mbox{Re}\;t_l^I(s)=q^{2l}\{a_l^I +q^2 b_l^I +O(q^4)\}~,\label{eqthrex}
\eeq
with $q$ the center--of--mass three--momentum of the pions, i.e.
$s = 4 (1 + q^2/M_\pi^2)$.
The threshold parameters $a_l^I$, $b_l^I$ are referred to as scattering lengths
and slope parameters, respectively.

In table \ref{tab:thresh}, we compare the threshold parameters of the low
partial waves with experimental
results. We emphasize once more that we do not consider the values
for either set I or set II as the definitive results of $O(p^6)$.
To arrive at such results, a
more detailed analysis based on the Roy--equation approach
is under way \cite{royanalysis}. With this caveat in mind, we notice that the
$\olc{l}{_1}$, $\olc{l}{_2}$ from the $D$--wave scattering lengths (set II)
tend to
improve the agreement with experiment with the possible exception of the
$S$--wave scattering lengths $a_0^0$, $a_0^2$. We comment on the experimental
determinations of scattering lengths below. We also note
that the ambiguity in scale or in the choice of
the lagrangian of $O(p^4)$ affects $a_0^0$ in the third digit only.

\begin{table}[t]
\caption{Threshold parameters in units of $M_{\pi^+}$. The values of $O(p^4)$
and $O(p^6)$ (set I) correspond to $\olc{l}{_1}=-1.7$, $\olc{l}{_2}=6.1$,
$\olc{l}{_3}=2.9$, $\olc{l}{_4}=4.3$. For $O(p^6)$ (set II),
$\olc{l}{_1}=-1.5$,
$\olc{l}{_2}=4.5$ were extracted from the $D$--wave scattering lengths
($\olc{l}{_3}$, $\olc{l}{_4}$ unchanged). The fourth column shows the values
obtained taking only the contributions from the $k_i$
\protect\cite{colangelo} defined in App. D.
We set $\mu = 1~\rm{GeV}$ and take
the $r_i^r(1~\rm{GeV})$ from (\protect\ref{eqrir}). The experimental
values are from Ref.~\protect\cite{threshexp}. }
\label{tab:thresh}

\vspace*{1cm}

\begin{center}
\begin{tabular}{c||c|c|c|c|c|c}
\hline
& & &\multicolumn{2}{c|}{} & & \\
  &\mbox{  } $O(p^2)$\mbox{  } & \mbox{  }$O(p^4)$\mbox{  }&
\multicolumn{2}{c|}{\mbox{  } $O(p^6)$ set I \mbox{  }}&
\mbox{  }$O(p^6)$\mbox{  } &\mbox{  } experiment\mbox{  } \\
 & & &\multicolumn{2}{c|}{$k_i$ \qquad all}& set II & \\[2pt]
\hline
$ a_0^0$& $0.16$ & $0.20$  &$0.213$& $0.217$ & $0.206$ & $0.26\pm0.05$\\

$b_0^0$ & $0.18$ & $0.25$&$0.279$& $0.275$ &$0.249$  & $0.25\pm0.03$\\

$-$10 $a_0^2$ & $0.45$ & $0.42$ &$0.407$&$0.413$ & $0.443$  & $0.28\pm0.12$\\

$-$10 $b_0^2$ & $0.91$ & $0.73$ & $0.69$&$0.72$ & $0.80$ & $0.82\pm0.08$\\

10 $a_1^1$    & $0.30$  & $0.37$  &$0.40$& $0.40$ & $0.38$ & $0.38\pm0.02$\\

$10^2 b_1^1$     & $0$ & $0.48$  &$0.78$& $0.79$ & $0.54$  & \\

$10^2 a_2^0$     & $0$ & $0.18$  &$0.30$ & $0.27$ & input & $0.17\pm0.03$ \\

$10^3 a_2^2$     & $0$ & $0.21$  &$0.34$ & $0.23$ & input & $0.13\pm0.30$ \\
\hline
\end{tabular}
\end{center}
\end{table}

\begin{figure}[t]
\begin{center}
\mbox{\epsfysize=10cm \epsfbox{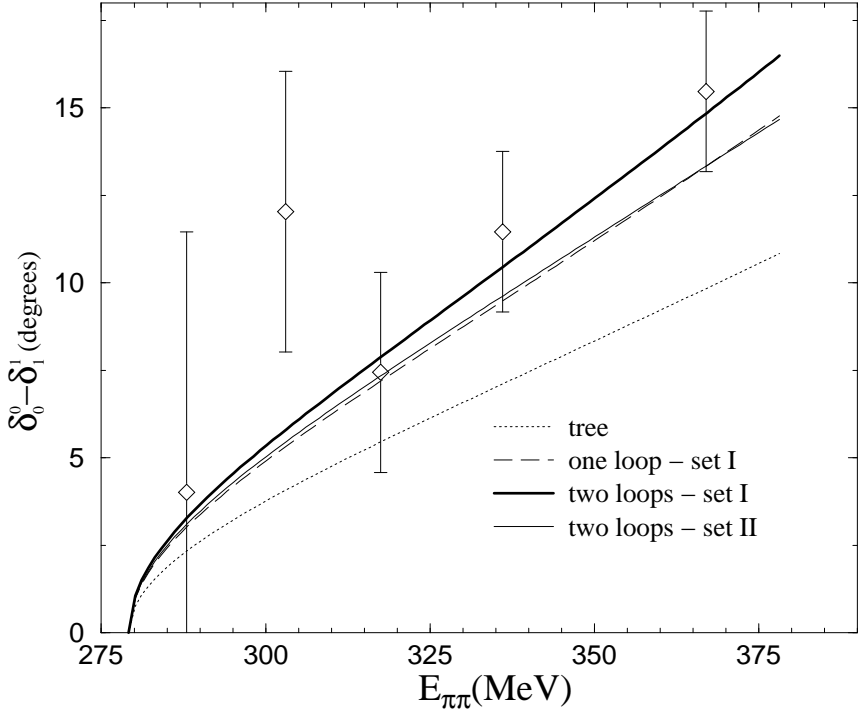}
     }
\caption{Phase shift difference $\delta_0^0-\delta_1^1$ at $O(p^2)$,
  $O(p^4)$ and $O(p^6)$ (set I and II). The data points are from
  Ref.~\protect\cite{rosselet}.}
\label{fig:d00md11}
\end{center}
\end{figure}

\begin{figure}[t]
\begin{center}
\mbox{\epsfysize=10cm \epsfbox{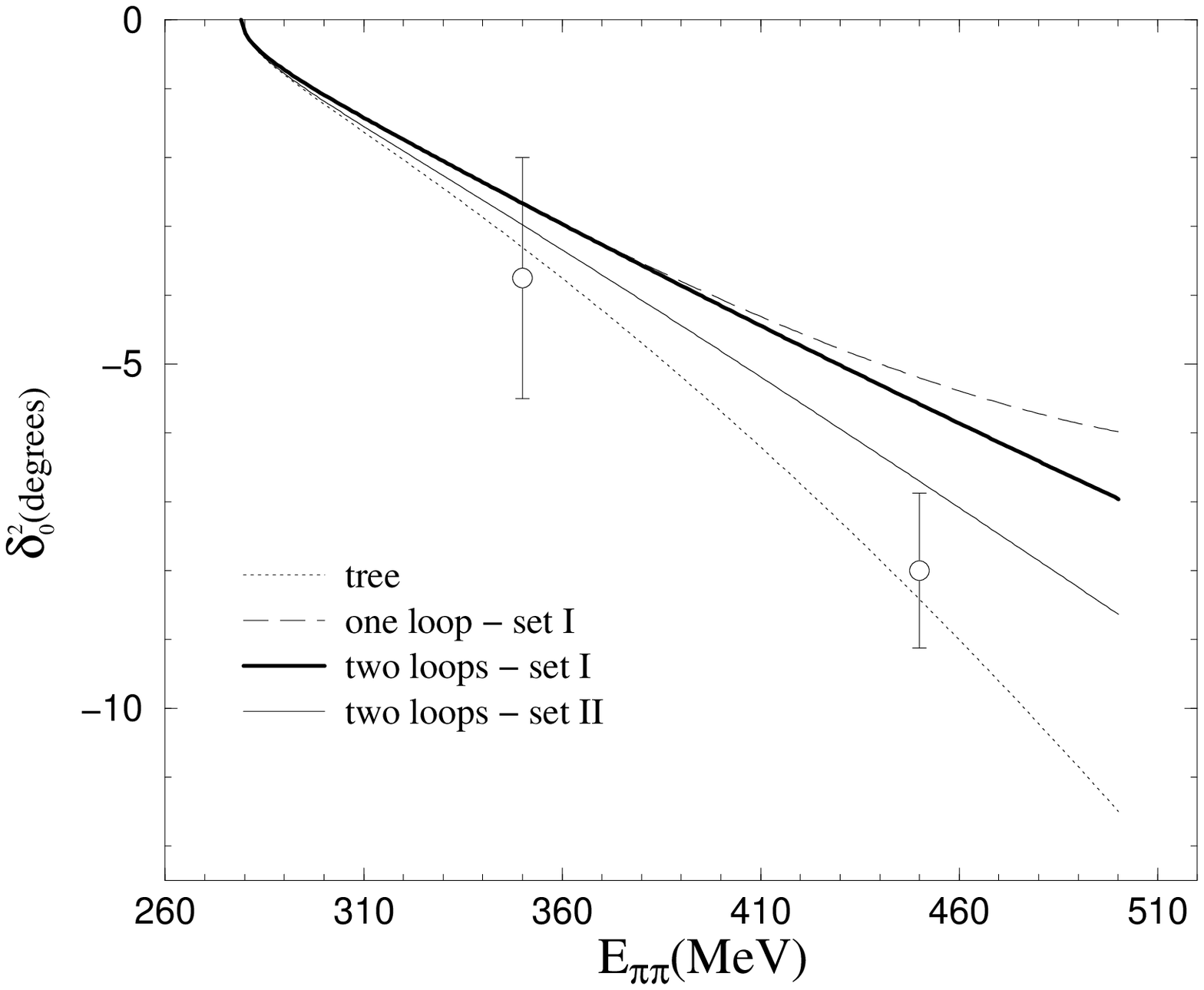}
     }
\caption{Phase shift $\delta_0^2$ at $O(p^2)$,
  $O(p^4)$ and $O(p^6)$ (set I and II). The data points are from
  Ref.~\protect\cite{hoogland}. }
\label{fig:d20}
\end{center}
\end{figure}

In Figs. \ref{fig:d00md11}, \ref{fig:d20} we plot the phase shift
difference $\delta_0^0 - \delta_1^1$
and the $I=2$ $S$--wave phase shift $\delta_0^2$ as functions of the
center--of--mass energy and compare with the available low--energy data.
The phase shifts correspond to the amplitudes of $O(p^2)$, $O(p^4)$ and
$O(p^6)$ (for both sets I and II), respectively. The two--loop phase shifts
describe the $K_{e4}$ data quite well for both sets of $\olc{l}{_i}$,
with a small
preference for set I. The $I=2$ $S$--wave, on the other hand, seems to prefer
set II.

There is an interesting lesson we can draw from our amplitude
concerning the extraction of threshold parameters from phase shift data.
Looking at Fig. \ref{fig:d00md11}, we observe that the $p^6$ amplitude
(set I) describes the data almost perfectly. On the other hand, the
value for the scattering length $a_0^0$ in table \ref{tab:thresh}
is 0.217 for set I,  quite a bit smaller than the 0.26 from
Ref.~\cite{threshexp} and even smaller than the 0.28 from
Ref.~\cite{rosselet}.

To put these differences into perspective, we consider the effective range
approximation for $S$--wave scattering \cite{collision},
\begin{equation}
q\,\cot{\delta_0^I} = \frac{M_\pi}{a_0^I} + \frac{1}{2}r_0^I q^2~,
\label{eqeffrange}
\end{equation}
with $r_0^I$ the effective range. In terms of the threshold
parameters defined in (\ref{eqthrex}), the effective range is given by
\begin{equation}
r_0^I = \frac{1}{M_\pi a_0^I}-\frac{2 M_\pi b_0^I}{(a_0^I)^2}-
\frac{2 a_0^I}{M_\pi} \,.
\end{equation}
A similar formula can be used for $\delta_1^I$ when we replace
$(a_0^I,b_0^I)$ by $(q^2 a_1^I, q^2 b_1^I)$.

In Ref.~\cite{rosselet}, the following approximation was used to extract
the scattering length $a_0^0$ from the measured phase shifts:
\beq
\sin{2(\delta_0^0-\delta_1^1)} = 2 \sqrt{1-\dfrac{4}{s}}\; (a_0^0 +
q^2 b)~.\label{eqross}
\eeq
In addition, a relation between $a_0^0$ and $b$ attributed to
Basdevant et al.~\cite{basdevant} was employed for a one--parameter fit to
the data.

\begin{figure}[t]
\begin{center}
\mbox{\epsfysize=10cm \epsfbox{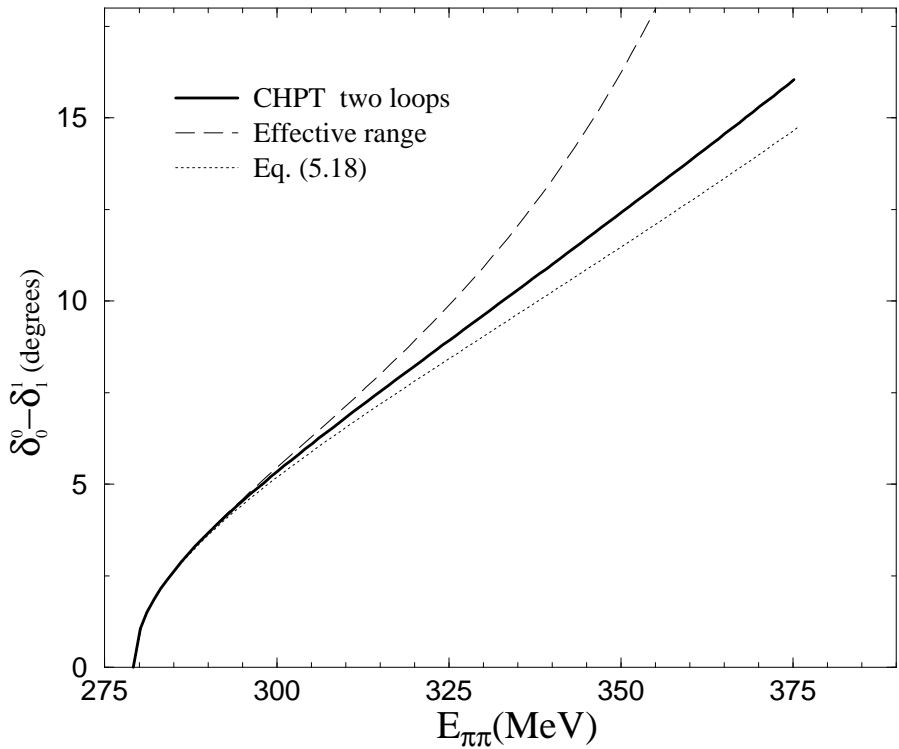}
     }
\caption{Comparison of the full two--loop phase shift difference
  $\delta_0^0-\delta_1^1$ with the
  effective range formula (\protect\ref{eqeffrange}) and with
  Eq.~(\protect\ref{eqross}). }
\label{fig:effr}
\end{center}
\end{figure}

In Fig. \ref{fig:effr}, we compare the phase shift difference $\delta_0^0$ -
$\delta_1^1$
{}from the full two--loop calculation (set I) with
the effective range formula (\ref{eqeffrange}) and with the approximation
(\ref{eqross}) used by Rosselet et al.~\cite{rosselet}. For the
effective range approximation, we have used the threshold parameters
corresponding to the CHPT amplitude (set I in table \ref{tab:thresh}).
In the curve based on (\ref{eqross}) we have also used $a_0^0=0.217$ and
the above mentioned relation \cite{basdevant} between $a_0^0$ and $b$.

The obvious conclusion is that both the effective range approximation and
the approximation used in Ref.~\cite{rosselet} deviate from
CHPT to $O(p^6)$ already at comparatively low
energies. Fig. \ref{fig:effr} also shows
why the scattering length obtained in Ref.~\cite{rosselet}
is bigger than the CHPT value.
It is hardly necessary to emphasize that the phase shift from CHPT is
superior to both approximations on all accounts.

\setcounter{equation}{0}
\section{Conclusions}

In this paper we have presented the calculation of elastic pion--pion
scattering to
sixth order in the low--energy expansion of QCD. The main part has been
devoted to explaining the technical aspects of the results given
in \cite{BCEGS1}.

We first developed the loop expansion and the renormalization procedure
for the generating functional of Green functions at the one-- and two--loop
level in the
$N$--component $\phi^4$ theory
We concentrated on issues that are relevant for the corresponding CHPT
calculation. In particular, we have discussed the dependence
on the renormalization scheme and the impact of the EOM.
Using the EOM, the counterterms in $\phi^4$ theory can be written in different
ways. We demonstrated explicitly that different choices are equivalent in the
sense
that the differences can always be absorbed into the higher--order parameters.
We discussed a method for calculating two--loop diagrams by treating the
renormalized one--loop amplitude as a nonlocal vertex. This approach turns
out to be especially useful for the  more involved CHPT calculation.

For the case of CHPT, we then performed the corresponding calculations for
the pion mass, the pion decay constant and the $\pi\pi$ scattering amplitude.
The renormalization of these quantities to $O(p^6)$ was discussed in detail
for both minimal subtraction and for the modified minimal subtraction scheme
that we actually used. In analogy to $\phi^4$ theory, the role of the EOM
in relating different forms of the chiral lagrangian was discussed. We also
commented on the Laurent expansion in $d - 4$ of the coupling constants of
$O(p^4)$.

We determined the complete dependence of $M_\pi$, $F_\pi$ and the scattering
amplitude on the quark masses.
As in previous full two--loop calculations \cite{BGS94,Bur96,BT97},
this dependence is both of theoretical interest and of numerical relevance.
The analytical expressions for the scattering amplitude and for the threshold
parameters were given.

For the numerical analysis of $\pi\pi$ scattering, we derived
estimates of the low--energy constants of $O(p^6)$ on the basis of meson
resonance exchange including kaon and eta contributions. These estimates
turn out to be dominated by vector meson contributions. Although there
are considerable uncertainties in this simple version of resonance
saturation, the overall size of the constants of $O(p^6)$ is such that many
quantities are relatively immune to those uncertainties. This is especially
the case for the $S$--waves. However, for the higher partial waves and for a
systematic error analysis in general a more sophisticated approach is needed.
Such an analysis based on Roy equations is under way \cite{royanalysis}.
A similar approach has already been used in the dispersive treatment of the
scattering amplitude to two--loop order \cite{KMSF95,KMSF96}.

The major uncertainties for a numerical analysis reside in the low--energy
constants of $O(p^4)$. To illustrate this uncertainty, we have presented
results both for the standard set of those constants \cite{glann,bcgke4} and
for a second set where $\olc{l}{_1}$, $\olc{l}{_2}$ are determined
from the $D$--wave threshold parameters to $O(p^6)$ as given in this paper.
In this case, even the $S$--waves are sensitive
to which set of constants is used.
This makes it all the more necessary to perform an analysis
where all the constants are determined on the basis of $O(p^6)$ quantities that
include, in particular, quark mass effects.

We have also presented numerical results for the phase shifts including the
combination that can be measured directly in $K_{e4}$ decays. For this
particular example, we demonstrated that approximations like the effective
range expansion may deviate from the CHPT predictions already at relatively low
energies. We have emphasized the problems of extracting threshold parameters
on the basis of such approximations from phase shift data.

The results obtained here for the $\pi\pi$ scattering amplitude to $O(p^6)$
will be used for a systematic analysis \cite{royanalysis} of the available
experimental data. In addition, they will be the basis for significant tests
of QCD at low energies together with forthcoming measurements of $\pi\pi$
scattering
near threshold in $K_{e4}$ decays and in pionium decay \cite{nemenov}.
\vskip.4cm
\noindent {\bf Acknowledgements}

We would like to thank M. Knecht, H. Leutwyler and J. Stern for useful
discussions.
This work is supported in part
by  FWF (Austria), Project No. P09505--MAT,
by the Magnus Ehrnrooth Foundation
(Finland), by the Swedish Science Research Council (NFR),
by the Swiss National Science Foundation,
by the Bundes\-amt f\"ur Bildung und Wissenschaft (contract No.
930341),
 and by HCM, contract No. CHRX--CT920026 (EURODA$\Phi$NE).

\appendix
\setcounter{equation}{0}
\newcounter{zahler}
\addtocounter{zahler}{1}
\renewcommand{\thesection}{\Alph{zahler}}
\renewcommand{\theequation}{\Alph{zahler}.\arabic{equation}}
\section{Fish and sunset diagrams}
\label{apdiagrams}
\subsection{Notation}
We use the notation
\bea
\la\ldots\ra = \int \frac{d^dl_1}{i(2\pi)^d}(\ldots)\scs
\la\la\ldots\ra\ra=\int\frac{d^dl_1}{i(2\pi)^d}\int\frac{d^dl_2
} {i(2\pi)^d}(\ldots)\co\nonumber
\eea
together with
\bea
w=\frac{d}{2}-2\per\pnn
\eea
Furthermore, as only one mass parameter occurs in this
appendix, we put everywhere $M^2=1$. It is straightforward to supplement
the quantities below with the relevant mass factors.

 The loop function $J(s)$ in
$d$ dimensions is \bea\label{eqfj}
J(s)&=&\bla \frac{1}{(1-l_1^2)(1-(l_1-p)^2)}\bra\nn
&=& \frac{1}{(4\pi)^{2+w}}\Gamma(-w)\int_0^1dx\,
(1-sx(1-x))^w\;\; ; \;\; p^2=s\per
 \eea
For ${J}$ one has  the dispersive
representation
 \bea\label{eqcauchy}
J({s})=\int_4^\infty\frac{[dx]}{x-{s}}\;\; ;\;\;
-1.5<w<0\scs
\eea
where the measure is
\bea
[dx]&=&\frac{\Gamma(3/2)}
{(4\pi)^{2+w}\Gamma(3/2 + w)}\left(\frac{x}{4}\right)^w
 (1-\frac{x}{4})^{\frac{1}{2}+w}
\;dx\per\label{eqmeasure}
 \eea
It is often convenient to split off the divergent part
through
 \bea\label{eqdecj}
J(s)&=&J(0)+\bar{J}(s)\scs\nn
J(0)&=&
 \frac{1}{(4\pi)^{2+w}}\Gamma(-w)\scs\nn
\bar{J}(s)&=&s
\int_4^\infty\frac{[dx]}{x(x-{s})}\nn
&\stackrel{d\rightarrow
 4}{=}&\frac{1}{16\pi^2}
\left[\sigma \ln \frac{\sigma-1}{\sigma+1}+2\right]
\;\;\; ; \; \;\;
\sigma=\left(1-4/{s}\right)^\frac{1}{2}\per
\eea
The constant $J(0)$ is related to $\dot{T}_M$, used in the text, via
$\dot{T}_M= -J(0)$ taking into account that $M=1$ in this appendix.

\subsection{The fish diagram}
The scalar integral for the fish diagram has the form
\bdm
V(s)=\bla\bla\,
\prod_{i=1}^4\frac{1}{D_i}\bra\bra
 \edm
with
\bea
D_1&=&1-l_1^2\scs D_2=1-(Q-l_1)^2\scs\nn
D_3&=&1-l_2^2\scs D_4=1-(l_2+l_1-p_1)^2\scs\nn
Q&=&p_1+p_2\scs p_1^2=p_2^2=1\scs Q^2=s\per\pnn
\eea
We evaluate this integral in two steps: First, we identify that
part of the function $V(s)$ that stays finite as $d\rightarrow
4$ by subtracting the divergent subintegral and by removing the
remaining overall divergence. In the second step, we
determine the finite part by evaluating its
absorptive part along the  cut in the variable $s$ and
by then constructing an
analytic function that has the same absorptive part and the same
 behaviour at $s\rightarrow 0,-\infty $.\\
\vskip.2cm
{\underline{Subtractions}}\\
\vskip.2cm
Integration over $l_2$ generates the loop function $J(\bar{t}),
\bar{t}=(p_1-l_1)^2$.
 We subtract the  divergence in this subdiagram by using the
decomposition (\ref{eqdecj})
 and obtain \bea
V(s)&=&J(s)J(0)+V_1(s)\scs\nn
V_1(s)&=&
\int_4^\infty
\frac{[dx]}{x} \bla\frac{
\bar{t}}{D_1D_2(x-\bar{t})}\,
\bra\,.
\eea
$V_1$ is still divergent at $d=4$. Using the Feynman parametrization
\bdm
\frac{1}{D_1D_2(x-\bar{t})}=\int_0^1\frac{2x_2dx_1dx_2}
{\left[\left(x_1D_1+(1-x_1)D_2\right)x_2+\left(x-\bar{t}\right)
\left(1-x_2\right)\right]^3}\scs
\edm
integrating first over $l_1$ and then  by parts in
$x_1$
shows that $V_1$ can be made finite by subtracting its value
at $s=0$,
\bdm
V_1(s)=V_1(0)+\overline{V}_1(s)\;\; ;\;\;
\lim_{s\rightarrow -\infty}\frac{\overline{V}_1}{s}
=0\per
\edm
$V_1(0)$ can be evaluated by methods similar to the ones used
below for the sunset integral.\\
\vskip.2cm
{\underline{ The finite part
$\overline{V}_1(s)$}}\nopagebreak
 \vskip.2cm
To evaluate the absorptive part of $\overline{V}_1(s)$, we invoke
unitarity \cite{nakanishi}:
\bea
\mbox{Im} V(s)&=&J(0) \mbox{Im} J(s) +\mbox{Im} \overline{V}_1(s)\nn
&=&\frac{(2\pi)^{2w+4}}{2}\int
d\mu(l)d\mu(l')\delta^{(d)}(Q-l-l')J(\bar{t})\co\pnn
\eea
where $d\mu$ is the Lorentz invariant measure in $d$ dimensions,
\bea
d\mu(l)=\frac{d^{2w+3}l}{2(2\pi)^{2w+3}l^0}\scs
l^0=\sqrt{1+{\vec{l}}^2}\scs\label{eqlinvm}\per
\eea
We write again $J(\bar{t})=J(0)+\bar{J}(\bar{t})$ and use
\bea
\frac{(2\pi)^{2w+4}}{2}\int
d\mu(l)d\mu(l')\delta^{(d)}(Q-l-l')&=&\nn
&&\hspace{-3cm} \frac{\pi\Gamma(3/2)}
{(4\pi)^{2+w}\Gamma(3/2 + w)}\left(\frac{s}{4}\right)^w
 \left(1-\frac{s}{4}\right)^{\frac{1}{2}+w}\per\pnn
\eea
This expression agrees with the absorptive part of $J(s)$ in
$d$ dimensions, see Eqs. (\ref{eqcauchy}, \ref{eqmeasure}).
Therefore, at $d=4$, one has
\bea
\mbox{Im} \overline{V}_1(s)=\frac{(2\pi)^4}{2}
\int d\mu(l)d\mu(l')\delta^{(4)}(Q-l-l')\bar{J}(\bar{t})\scs
\label{eqimvbar}
\eea
where the measure in (\ref{eqimvbar}) is now the ordinary
four--dimensional
one, obtained from (\ref{eqlinvm}) by putting $w=0$. For
$\bar{J}$ we  insert its explicit expression (\ref{eqdecj}) at
$d=4$. It is convenient to work
out the  phase space integral (\ref{eqimvbar}) in the center--of--mass
frame where
 \bdm
Q^\mu=(\sqrt{s},\vec{0})\scs
 p_1^\mu=(\sqrt{s}/2,0,0,p)\scs p=\sqrt{s/4-1}\per
\pnn
 \edm
After integration over $l'$, one is left with an integral over
the three--momentum $\vec{l}$.
 Using the remaining one--dimensional delta function,
all integrals except the one over the angle $\theta
(\vec{l},\vec{p}_1)$
can be done easily. The integral left over is proportional to
\bdm
\int_{-1}^{1}dz\left[\sigma_1
\ln\frac{\sigma_1-1}{\sigma_1+1}+2\right]\scs
\sigma_1=\sqrt{1+\frac{2}{p(1-z)}}\per\pnn
 \edm
After the change of variables
$(\sigma_1 -1)/(\sigma_1+1)=v$, we find
\bea
\mbox{Im}\,\overline V_1(s) = \frac{1}{4 (4 \pi)^3} \left[ 3
\sigma + \ln \frac
{1-\sigma}{1+\sigma}- \frac{1}{s \sigma} \ln^2 \frac{1-\sigma}{1+\sigma}
\right]\,,\;\; \sigma=\left(1-4/s\right)^\frac{1}{2}\per
\label{A8}
\eea
It remains to construct a function with the proper cut structure
and the correct behaviour at $s\rightarrow 0,\infty$. We find
\bea
\overline{V}_1(s)=
 \frac{1}{(16 \pi^2)^2} \left[
\left( 3-\frac{\pi^2}{3 s \sigma^2}\right) f +\frac{1}{2
\sigma^2} f^2-
\frac{1}{3 s \sigma^4} f^3  +6 + \frac{\pi^2}{6} \right]
\pnn
\eea
 with
\bea
f= \sigma \ln \frac{1-\sigma}{1+\sigma} + i \pi \sigma.\pnn
\eea
In the notation of Ref. \cite{BCEGS1}, this is
\bea
\overline{V}_1(s)=\frac{3}{N}\bar{J}(s)+\frac{{K}_1(s)}{2}
                  -\frac{{K}_3(s)}{3}\per\pnn
\eea
Of course, exactly the same method can be applied to  integrals
with a more complicated numerator.

\subsection{The sunset diagram}

For the sunset diagram integrals of the type
\bea\label{eqa14}
(H;H^{\mu};H^{\mu \nu}) = \bla\bla
\frac{(1;l^{\mu}_1;l^{\mu}_1l^{\nu}_1)}
{[1-l_1^2] [1-l_2^2] [1-(p-l_1-l_2)^2]}\bra\bra
\eea
have to be calculated. Here we again focus on the scalar integral $H$, the
remaining ones can be done analogously.
For $H$ we also present the procedure for the infinite part.
Using the $d$-dimensional dispersion representation
(\ref{eqcauchy}) for $J$,  we get
\bea
H = \int_4^\infty [ds'] \; \int_0^1 dx\,
\bla\frac{1}{[x+s'(1-x)-p^2
x(1-x)-l_1^2]^2}\bra\per\pnn \eea
The $d^dl_1$ integrals can be performed easily and the result is
\bea
H(p^2) = \int_4^{\infty} [ds'] \int_0^1 dx\, F_2[z_2]
\label{eqa15}
\eea
where
\bea
F_2[z_2] &=& \frac{\Gamma(-\omega)}{(4 \pi)^{2+\omega}} \;
z_2^{\omega} \scs\nn
z_2(p^2) &=& x+s' (1-x)-p^2 x (1-x).
\eea
We subtract and add the two first terms of the Taylor series of
$F_2[z_2]$ in $p^2$ around $p^2=1$ and obtain for the finite part
at $d=4$
\bea\label{eqa16}
H(p^2)-H(1)-(p^2-1)H'(1)
=\int_4^{\infty} ds' \;
\sqrt{1-\frac{4}{s'}} \int_0^1 dx \; {\cal K}_2(s',x;p^2), \pnn
\eea
where we have introduced the kernel
\bea
{\cal{K}}_2(s',x;p^2)&=&\frac{1}{16\pi^2}\lim_{w\rightarrow
0}\left\{
F_2[z_2(p^2)]-F_2[z_2(1)]-(p^2-1){F}'_2[z_2(1)]\right\}\nn
&=&-\frac{1}{(16\pi^2)^2}\left\{\ln\frac{z_2(p^2)}{z_2(
1)}+\frac{(p^2-1)x(1-x)}{z_2(1)}\right\}.\pnn
\eea
The integral $\int\,dx\,{\cal K}_2$ could further be done in closed form
-- the result amounts to a twice subtracted one--loop self--energy
integral with two propagators (with masses $1$ and $s'$).
In the text, we  use
\bea
\ol{\ol{H}}(p^2)
=\frac{1}{(p^2-1)^2}\int_4^{\infty} ds' \;
\sqrt{1-\frac{4}{s'}} \int_0^1 dx \; {\cal K}_2(s',x;p^2)\co\label{eqa50}
\eea
which is finite  at $p^2=1$.
We also need
$H(1)$ and $H'(1)$, where the poles at $\omega=0$ are contained.
For the evaluation of the infinite parts a recursion relation can be set up
by partial integration in $x$
in (\ref{eqa15}). This method gives \cite{BGS97}
\bea
H(1)&=&-\frac{1}{(4 \pi)^{4+2\omega}}\Gamma^2(-\omega)\left\{\frac{3}{2}
-\frac{17}{4}\omega+\frac{59}{8}\omega^2+O(\omega^3)\right\}\co\nn
H'(1)&=&-\frac{2}{(4 \pi)^{4+2\omega}}\Gamma^2(-\omega)\left\{\frac{1}{8}
\omega+\frac{3}{16}\omega^2+O(\omega^3) \right\}.
\label{eqa51}
\eea
In Ref.~\cite{post}, the evaluation of the sunset integral is
discussed in the general mass case.

\setcounter{section}{0}
\setcounter{subsection}{0}

\renewcommand{\thesection}{\Alph{zahler}}
\renewcommand{\theequation}{\Alph{zahler}.\arabic{equation}}

\setcounter{equation}{0}
\addtocounter{zahler}{1}
\renewcommand{\thesection}{\Alph{zahler}}
\renewcommand{\theequation}{\Alph{zahler}.\arabic{equation}}

\section{Off--shell four--point function in $d$ dimensions}
\label{appipi}
We display the four--point function
 to one
loop in $d$ dimensions. More precisely, we use the sigma model
parametrization (\ref{eqsigma}) and define
\bea
&&i^3\int dx_1\,dx_2\,dx_3\,
e^{-i(p_1x_1+p_2x_2-p_3x_3-p_4x_4)}\langle0|T\phi^i(x_1)
\phi^k(x_2)\phi^l(x_3)\phi^m(x_4)|0\rangle\nn
&&=\frac{Z^2}{\prod_i(M_\pi^2-p_i^2)}T^{lm;ik}(s,t,u;p_1^2
, p _2^2,p_3^2,p_4^2)\nonumber
\eea
with\footnote{In Sect. 4, we express the Mandelstam variables
in units of the physical pion mass squared. For simplicity of
notation, we use in this appendix the standard definition.}
 \bea
&&p_1+p_2=p_3+p_4\co\nn
&&s=(p_1+p_2)^2\scs t=(p_1-p_3)^2\scs u=(p_1-p_4)^2\per\pnn
\eea
The wave function renormalization constant $Z$ is the one
appearing in the two--point function (\ref{eqzphi}). In the
standard isospin
decomposition \bea T^{lm;ik}(s,t,u;p_1^2,p_2^2,p_3^2,p_4^2)=
\delta^{ik}\delta^{lm}A(s,t,u;p_1^2,p_2^2,p_3^2,p_4^2)
+{\mbox{ cycl.}}
\co\nonumber
\eea
the scattering amplitude is obtained by putting all momenta on
the mass shell,
\bea
A(s,t,u)=A(s,t,u;M_\pi^2,M_\pi^2,M_\pi^2,M_\pi^2)=\frac{s-M^2}
{F^2} + O(p^4)\per\pnn
\eea
We find
\bea
F^4A(s,t,u;p_1^2,p_2^2,p_3^2,p_4^2) &=&\nn
&&\hspace{-2cm} (s-M^2)\left\{F^2+
 (2 s +t+u-3 M^2) J_1(s)\right\}
        \nn
 & &\hspace{-2cm}
        +\left\{\left[p_1^2 p_4^2+p_2^2 p_3^2-t
(p_2^2+p_3^2-t)\right]J_1(t)
           \ \right. \nn
 & &\left. \hspace{-2cm}
        + \left[\triangle_{13}\triangle_{24}
-t(\triangle_{13}-\triangle_{24}+t)\right]J_2(t)          \right.
\nn  & &\left. \hspace{-2cm}
          -2t\left[p_1^2+p_4^2-u\right]J_3(t)
+(p_3,p_4,t)\rightarrow (p_4,p_3,u)\right\}\nn
&&\hspace{-2cm}
+8l_1p_1p_2\cdot p_3p_4
+4l_2\left[p_1p_3\cdot p_2p_4+p_1p_4\cdot p_2p_3\right] \nn
&&\hspace{-2cm}+
\left[s+p_2^2-p_1^2-5M^2/2\right]T_M
\co\nn \eea
where
\bea
J_1(s)&=&\frac{1}{2}J(s)\scs
J_2(s) = -\frac{1}{s (d-1)}\left[\left(M^2-\frac{d}{4}s\right) J(s)
                +\left(\frac{d}{2}-1 \right)T_M \right]\co \nn
J_3(s) &=& \frac{1}{s (d-1)}\left[\left(M^2-\frac{s}{4}\right) J(s)
                - \frac{1}{2} T_M \right]\,\mbox{ and }\,
\triangle_{ik}=p_i^2-p_k^2\per\pnn\nn
\eea
\underline{Remarks:}\\
\begin{itemize}
\item[i)]
For on--shell momenta, the result agrees at $d=4$ with the result
given in \cite{glann}.
\item[ii)]
For off--shell pions, the amplitude is not finite at $d=4$, in
contrast to the off--shell amplitude considered in \cite{glann},
where the four--point function of pseudoscalar densities
was considered. In that case, there are additional contributions
proportional to the low--energy constants $l_3$ and $l_4$ that remove the
remaining divergences.
\item[iii)]
It is not surprising that the above amplitude is not finite
off--shell: The construction given in Ref. \cite{glann} only
guarantees that the Green functions built from quark currents
are ultraviolet finite -- Green functions of pion fields are
unphysical objects, even if they occur at intermediate steps of a
calculation, as in the present context.
\item[iv)]
Finally, we mention that these divergences in the
off--shell
amplitude do not generate nonlocal singularities in the two--loop
calculation.
\end{itemize}
\addtocounter{zahler}{1}
\setcounter{equation}{0}
\section{Scattering lengths and effective ranges}
\label{apaik}
\newcommand{\bb}{\,\bar{b}}
{}From the explicit expression for the scattering amplitude in
equation (\ref{amptot1}), it is straightforward to evaluate the
threshold parameters $a^I_l$ and $b_l^I$.
Using the definition (\ref{eqthrex}), we find
\bea
a_0^0&=&\frac{7M_\pi^2
}{32\pi
F_\pi^2}\left\{1+\frac{x}{7}\left[49+5\bb_1+12\bb_2+48\bb_3
+32\bb_4\right]\right.\nn
{}\nn
&&\left.+ x^2\left[\frac{7045}{63}-\frac{215\pi^2}{126}
+ 1 0 \bb_1
+24\bb_2+96\bb_3+64\bb_4+\frac{192}{7}\bb_5\right]\right\}
\co\nn
{} \nn
{} \nn
b_0^0&=&\frac{1}{4\pi F_\pi^2}
\left\{1+\frac{x}{4}\left[\frac{281}{9}+4\bb_2+48(\bb_3+\bb_4)
\right]\right.\nn
{}\nn
&&\hspace{-.8cm}\left.
+\frac{x^2}{4}\left[\frac{77489}{81}-\frac{4135\pi^2}{72}+
\frac{10
}{3}\bb_1+\frac{592}{9}\bb_2+\frac{6448}{9}\bb_3+688\bb_4+288\bb_
5-32\bb_6\right]\right\}\co\nn
{}\nn
{}\nn
 a_0^2&=&-\frac{M_\pi^2}{16\pi
F_\pi^2}\left\{1-x\left[2+\bb_1+16\bb_4
\right] +x^2\left[\frac{262}{9}-\frac{22\pi^2}{9}+4\bb_1
+64\bb_4\right]\right\}\co\nn
{}\nn
{}\nn
b_0^2&=&-\frac{1}{8\pi
F_\pi^2}\left\{1-\frac{x}{2}\left[\frac{97}{18}-2\bb_2
+ 48 \bb_4\right]\right.\nn
{}\nn
&&\left.+\frac{x^2}{2}\left[\frac{10591}{81}-\frac{145\pi^2}{12}
+\frac{11}{3}\bb_1-\frac{64}{9}\bb_2+\frac{32}{9}\bb_3+\frac{752}
{ 3}\bb_4+32\bb_6\right]\right\}\co\nn
{}\nn
{}\nn
a_1^1&=&\frac{1}{24\pi F_\pi^2}
\left\{1+x\left[-\frac{17}{36}+\bb_2+ 8\bb_4
\right]\right.\nn
{}\nn
&& \left.
+x^2\left[-\frac{181}{162}+\frac{7\pi^2}{24}-\frac{5}{6}\bb_1
-\frac{16}{9}\bb_2-\frac{16}{3}\bb_3-\frac{8}{3}\bb_4+16\bb_6
\right]\right\}\co\nn
{}\nn
{}\nn
b_1^1&=&\frac{1}{256\pi^3
F_\pi^4}\left\{\frac{37}{135}-\frac{8}{3}\bb_3+8\bb_4 + \right.
\nn {}\nn
&&\left.+x\left[\frac{56981}{2430}-\frac{337\pi^2}{810}-2\bb_1
-\frac{196}{135}
\bb_2-\frac{1888}{135}\bb_3-\frac{544}{135}\bb_4+\frac{64}{3}\bb_6
\right]\right\}\co\nn
{}\nn{}
{}\nn{}
a_2^0&=&\frac{1}{480 \pi^3 F_\pi^4}
\left\{-\frac{47}{72}+\bb_3+7\bb_4\right.\nn
{}\nn
&&\left.
+x\left[\frac{7003}{2160}+\frac{169\pi^2}{2160}+\frac{1}{1
2}\bb_1-\frac{11}{9}\bb_2-\frac{152}{45}\bb_3-\frac{364}{45}\bb_4
+32\bb_6\right]\right\}\co\nn
{}\nn{}
{}\nn{}
a_2^2&=&\frac{1}{480 \pi^3 F_\pi^4}
\left\{-\frac{49}{360}+\bb_3+\bb_4\right.\nn
{}\nn
&&\left.
+x\left[-\frac{67}{2160}-\frac{127\pi^2}{432}+\frac{29}{60}\bb_1+
\frac{19}{90}\bb_2+\frac{28}{45}\bb_3+\frac{188}{45}\bb_4+8\bb_6
\right]\right\}\co\nn
  \eea
 where \bea
 x=\frac{M_\pi^2}{16\pi^2F_\pi^2}\scs
\bb_{1,2,3,4}=16\pi^2b_{1,2,3,4}\,\mbox{ and }\,
\bb_{5,6}=(16\pi^2)^2b_{5,6}\per\nn
\eea
The coefficients $b_i$ are displayed in appendix \ref{apbi}.
The terms between the last square brackets
in the expressions for $a_l^I$ and $b^I_l$ generate contributions of
$O(p^8)$ --  these are beyond the accuracy we aim at here.
In order to keep the formulae
as simple as possible, we nevertheless retain them. In our numerical results
these are removed.

\addtocounter{zahler}{1}
\setcounter{equation}{0}
\section{The constants $b_i$}
\label{apbi}
The quantities  $b_i$ in Eqs. (\ref{amptot}) and
(\ref{amptot1}) stand for
\begin{eqnarray*}
b_1 &=& 8 \,l_1^r+2 \,l_3^r-2 \,l_4^r+\frac{7}{6} L +
\frac{1}{16{ \pi}^{2}} \frac{13}{18} \\
 &+& { x_2}
 \left\{ {\vrule height0.86em width0em depth0.86em} \right. \! \!
\frac{1}{16{ \pi}^{2}} \left[ \frac{56}{9}\,l_1^r + \frac{80}{9}\,l_2^r
+15 \,l_3^r +\frac{26}{9}\,l_4^r+\frac{47}{108}L-\frac{17}{216}
+  \frac{1}{16{ \pi}^{2}} \frac{3509}{1296} \right]  \\
 & & \; \; \; \; \; \; \;
+ \frac{1}{6}\left[4k_1+ 28k_2-6 k_3+ 13 k_4 \right]
+ \left[32 \,l_1^r+12 \,l_3^r - 5 \,l_4^r \right]
\,l_4^r - 8 {\,l_3^r}^2
+ r_1^r
 \left. {\vrule height0.86em width0em depth0.86em} \right\} \co \\ \\
b_2 &=& -8 \,l_1^r + 2 \,l_4^r-\frac{2}{3}L
-\frac{1}{16{ \pi}^{2}}\frac{2}{9} \\
&+& { x_2}
 \left\{ {\vrule height0.86em width0em depth0.86em} \right. \! \!
\frac{1}{16{ \pi}^{2}}\left[-24 \,l_1^r-\frac{166}{9}\,l_2^r-18 \,l_3^r -
\frac{8}{9} \,l_4^r -\frac{203}{54}L + \frac{317}{3456}
-\frac{1}{16{ \pi}^{2}} \frac{1789}{432} \right] \\
 & & \; \; \; \; \; \; \; \; \; \; \; \;
-\frac{1}{6}\left[ 54 k_1+62 k_2 + 15 k_3 + 10 k_4 \right]
-\left[ 32 \,l_1^r+4 \,l_3^r -5 \,l_4^r \right] \,l_4^r +r_2^r
 \! \! \left. {\vrule
height0.86em width0em depth0.86em} \right\} \co \\ \\
b_3 &=& 2 \,l_1^r + \frac{1}{2}\,l_2^r -\frac{1}{2}L
-\frac{1}{16{ \pi}^{2}}\frac{7}{12}
+ { x_2}
 \left\{ {\vrule height0.86em width0em depth0.86em} \right. \! \!
\frac{1}{16{ \pi}^{2}}\left[
{\vrule height0.86em width0em depth0.86em} \right. \! \!
\frac{178}{9} \,l_1^r +
\frac{38}{3}\,l_2^r-\frac{7}{3} \,l_4^r -\frac{365}{216}L \\
 & & \; \; \; \; \; \; \; \;
- \frac{311}{6912} + \frac{1}{16{ \pi}^{2}}\frac{7063}{864}
\! \! \left. {\vrule height0.86em width0em depth0.86em} \right]
+2 \left[4\,l_1^r +\,l_2^r \right]\,l_4^r
+ \frac{1}{6}\left[38 k_1 + 30 k_2 -3 k_4 \right] + r_3^r
 \! \! \left.
{\vrule height0.86em width0em depth0.86em} \right\}\co \\ \\
b_4 &=& \frac{1}{2} \,l_2^r -\frac{1}{6}L
-\frac{1}{16{ \pi}^{2}}\frac{5}{36}
+ { x_2}
 \left\{ {\vrule height0.86em width0em depth0.86em} \right. \! \!
\frac{1}{16{ \pi}^{2}}\left[
{\vrule height0.86em width0em depth0.86em} \right. \! \!
\frac{10}{9}\,l_1^r + \frac{4}{9}\,l_2^r
-\frac{5}{9} \,l_4^r +\frac{47}{216} L \\
 & & \; \; \; \; \; \; \; \; \; \; \; \;
+\frac{17}{3456}  +\frac{1}{16{ \pi}^{2}} \frac{1655}{2592}
\! \! \left. {\vrule height0.86em width0em depth0.86em} \right]
+ 2 \,l_2^r \,l_4^r -\frac{1}{6}\left[k_1+4 k_2 + k_4 \right] + r_4^r
\! \left. {\vrule height0.86em width0em depth0.86em}
 \right\} \co \\ \\
b_5 &=& \frac{1}{16{ \pi}^{2}} \left[-\frac{31}{6} \,l_1^r -
\frac{145}{36} \,l_2^r + \frac{625}{288} L + \frac{7}{864}-
\frac{1}{16{ \pi}^{2}} \frac{66029}{20736} \right]\! -\!\frac{21}{16}k_1
\!-\!\frac{107}{96} k_2
+ { r_5^r} \co \\ \\
\end{eqnarray*}
\bea
b_6 &=&\frac{1}{16{ \pi}^{2}}
\left[-\frac{7}{18}\,l_1^r-\frac{35}{36}\,l_2^r + \frac{257}{864} L
+\frac{1}{432}-\frac{1}{16{ \pi}^{2}}\frac{11375}{20736} \right]
-\frac{5}{48}k_1 -\frac{25}{96}k_2
+ { r_6^r}\co\nn\label{eqb_i}
\eea
where
\bea
L&=&\frac{1}{16\pi^2}\ln{\frac{M_\pi^2}{\mu^2}}\scs\nn
k_i&=&(4\,l_i^r(\mu)-\gamma_i L)L \; ; \;
\gamma_1=\frac{1}{3}\scs
\gamma_2=\frac{2}{3}\scs
\gamma_3=-\frac{1}{2}\scs
\gamma_4=2\per
\eea
We have denoted by  $l_i^r$ ($r_i^r$) the renormalized, quark mass
independent couplings from ${\cal L}_4$ (${\cal L}_6)$, with \cite{glann}
$\displaystyle{\mu\frac{dl_i^r}{d\mu}=-\frac{\gamma_i}{16\pi^2}\per}$
In the text, we also use the parameters $\olc{l}_i$, defined by
\bea
l_i^r=\frac{\gamma_i}{32\pi^2}\left(\olc{l}_i+
\ln\frac{M_\pi^2}{\mu^2}\right)\per\label{eqbarli}
\eea
The scale dependence of $r_i^r$ is fixed by the requirement
$\displaystyle{\mu\frac{db_i}{d\mu}=0}$.

\end{document}